\documentclass[pss]{wiley2sp} % provides pss two-column style
\usepackage{amsmath}
\usepackage{bm}             

\usepackage{physics}
\usepackage[euler]{textgreek}
\usepackage[symbol]{footmisc}
\begin{document}
% Title of the article
\title{Operation of a Microfabricated Planar Ion-trap for Studies of a Yb$^{+}$-- Rb Hybrid Quantum System}

% Authors
\author{
Abasalt Bahrami\textsuperscript{\textsf{\bfseries 1},\Ast},
Matthias M\"uller\textsuperscript{\textsf{\bfseries 1}},
Martin Drechsler\textsuperscript{\textsf{\bfseries 1},\dag},
Jannis Joger\textsuperscript{\textsf{\bfseries 2}},
Rene Gerritsma\textsuperscript{\textsf{\bfseries 2}},
Ferdinand Schmidt-Kaler\textsuperscript{\textsf{\bfseries 1}}
}

% Present address: Departamento de F{\'i}sica, FCEyN, UBA and IFIBA, Conicet, Pabell{\'o}n 1, Ciudad Universitaria, 1428 Buenos Aires, Argentina

%E-mail-address of corresponding author
\mail{e-mail
  \textsf{abahrami@uni-mainz.de\\$^{\dag}$\,\,Present address: Departamento de F{\'i}sica, FCEyN, UBA and IFIBA, Conicet, Pabell{\'o}n 1, Ciudad Universitaria, 1428 Buenos Aires, Argentina.}}

% author's affiliations/addresses
\institute{%
  \textsuperscript{1}\, QUANTUM, Institut f{\"u}r Physik, Universit{\"a}t Mainz, Staudingerweg 7, 55128 Mainz, Germany\\
  \textsuperscript{2}\,Van der Waals-Zeeman Institute, Institute of Physics, University of Amsterdam, Science Park 904, 1098 XH Amsterdam, The Netherlands}

% Please select about four verbal keywords for your manuscript.
 %I don't know how many keywords should be put, put this can be changed anytime.

\abstract{\bf%
In order to study interactions of atomic ions with ultracold neutral atoms, it is important to have sub-$\mu$m control over positioning ion crystals. Serving for this purpose, we introduce a microfabricated planar ion trap featuring 21 DC electrodes. The ion trap is controlled by a home-made FPGA voltage source providing independently variable voltages to each of the DC electrodes. To assure stable positioning of ion crystals with respect to trapped neutral atoms, we integrate into the overall design a compact mirror magneto optical chip trap (mMOT) for cooling and confining neutral $^{87}$Rb atoms. The trapped atoms will be transferred into an also integrated chip-based Ioffe-Pritchard trap potential formed by a Z-shaped wire and an external bias magnetic field. We introduce the hybrid atom-ion chip, the microfabricated planar ion trap and use trapped ion crystals to determine ion lifetimes, trap frequencies, positioning ions and the accuracy of the compensation of micromotion.}

\maketitle   % please do not remove

% INTRODUCTION %%%%%%%%%%%%%%%%%%%%%%%%%%%%%%%%%%%%%%%%%%%%%%%%%%%%%%%%%%%%%%%%%%%%%%%%%%
%%%%%%%%%%%%%%%%%%%%%%%%%%%%%%%%%%%%%%%%%%%%%%%%%%%%%%%%%%%%%%%%%%%%%%%%%%%%%%%%%%%
\section{Introduction}

The development of laser cooling techniques to cool neutral atoms below $\mu$K temperatures and magnetic trapping has paved the way to explore quantum degeneracy in atomic gases \cite{Dalfovo.1999,Giorgini.2008} and to investigate a variety of quantum many-body phenomena \cite{Greiner.2003,Regal.2004}. On the other hand Paul traps use radio-frequency (RF) driven electric fields and static voltages (DC) to trap ions in a time-averaged trapping potential \cite{Leibfried.2003}. 

Recently, the interactions between neutral atoms and charged ions have been addressed for investigations of low temperature chemical reactions \cite{Hall.2011,Rellegert.2011,Ratschbacher.2012}, the study of polaron physics \cite{Cote.2002}, but also to allow for novel types of quantum simulation \cite{Bissbort.2013}. From the experimental side, the challenge is to combine both trapping techniques into a hybrid atom-ion trap to study the atom-ion system. At large inter-nuclear separations the collisions between ions and neutral atoms are dominated by an attractive polarization potential of the form 

\begin{equation}
\label{eq:interaction}
V=-\frac{\alpha_{p} q^2}{(4\pi\varepsilon_{0})^2r^4},
\end{equation}

where $r$ is the inter-nuclear separation, $q$ the charge of the ion, $\alpha_{p}$ the isotropic static electric dipole polarizability of the neutral atom and $\varepsilon_{0}$ the vacuum permittivity \cite{Cote.2000b}. In the s-wave scattering limit the characteristic length scale of the interaction of the potential is given by 

\begin{equation}
\label{eq:r}
r^{*}=\sqrt{\frac{\mu\alpha_{p}q^2}{(4\pi\varepsilon_{0}\hbar)^2}},
\end{equation}

where $\mu=m_{i}m_{a}/\left(m_{i}+m_{a}\right)$ is the reduced mass of the two particles involved in the collision and $\hbar$ is the reduced Planck constant \cite{Gribakin.1993}. At this separation of the particles the characteristic energy scale $E^{*}=\hbar^{2}/2\mu R^{*2}$ can be assigned. For the specific case of $^{87}$Rb atoms interacting with Yb$^{+}$ ions $r^{*}=307\thinspace$nm \cite{Miller.2000}. In contrast short-range interactions, e.g. for $^{87}$Rb atoms, require a wave packet overlap at distances of a few nm \cite{Chin.2010}. Typical processes emerging from two-body collisions of hetero-nuclear atoms are listed in Tab.~\ref{tab.atomion}. Among possible interaction processes, we are interested in  processes which are dominated by elastic collisions. The atom-ion scattering cross-section in elastic collisions $\sigma_{el} (E)$ is described by 

\begin{equation}
\label{eq:crosssection}
\sigma_{el} (E)=\pi\left(1+\frac{\pi^2}{16}\right)r^{*2}\left(\frac{\hbar^2}{\mu r^{*2}E}\right)^{1/3},
\end{equation}

where $E\approx k_{B}T_{\text{ion}}$ \cite{Cote.2000}. To resolve the interactions between the atoms and ions on the length scale of $r^*$, we need to have a sub-$\mu$m accuracy in positioning the ions. In order to position the trapped ions with such accuracy we have to use an ion trap which has enough DC electrodes to confine ions in different trapping regions and position them accurately. For this purpose we use a segmented planar ion trap with 21 DC electrodes which can be independently controlled. By applying DC voltages to each of the opposing DC electrodes ions can be transported all along the trap axis. To investigate the interactions between atoms and ions, we have to spatially overlap the atomic cloud and ion crystals in a single apparatus. A novel atom-ion trap introduced here combines a planar ion trap with an integrated atom trap into a single chip trap. This new chip design implies an infrastructural simplification of the setup featuring an optimal optical access and making the two traps more robust to temporal drifts of the positions of the two species to each other.

%This ion trap is operated in our experiment to trap  Yb$^{+}$ ions and we achieved ion positioing accuracy of $\pm2.5\mu$m which is limited to the pixelsize of EMCCD camera (8$\mu$m$\times$8$\mu$m) (Section Trap operation)

\begin{table}[t] % "h" puts figure or table at a position of code
  \caption{Possible processes during two-body collisions of a hetero-nuclear ion A$^{+}$ with an atom B at low energy. (1) The atom and ion pair may undergo an elastic collision. In case of a perfectly elastic collision the internal states of atom and ion do not change, (2) the pair might exchange spins. (3) An electron from B may hop to A$^{+}$ if the particles are close enough to each other. A charge transfer process between B and A$^{+}$ may take place in the following ways: radiative charge transfer, photoassociative charge transfer or non-radiative charge transfer. (4) An atom-ion pair in the excited continuum may decay spontaneously to a bound level of lower electronic state.}
\begin{tabular}[htbp]{@{}lll@{}}
    \hline
 &     Types of processes &\\
    \hline
   (1) & A$^{+}$+ B $\rightarrow$ A$^{+}$+ B  & Elastic collisions \cite{Cote.2000b}\\
   (2) & A$^{+}\ket{\downarrow}$ + B$\ket{\uparrow}\rightarrow$ A$^{+}\ket{\uparrow}$ + B$\ket{\downarrow}$  & Spin exchange \cite{Makarov.2003,Verhaar.1987} \\
   (3) & A$^{+}$+ B $\rightarrow$ A + B$^{+}$  & N-radiative charge transfer\\
   & A$^{+}$+ B $\rightarrow$ A + B$^{+}$+ $\gamma$  & Radiative \cite{Cooper.1984,Zygelman.1988,Zygelman.1989}\\
   &  A$^{+}$+ B $\rightarrow$ (AB)$^{+}$+ $\gamma$  & Photoassociative\\
  (4) &   A$^{+}$+ B $\rightarrow$ A$^{+}$B + $\gamma$  & Molecule formation \cite{Rakshit.2011}\\
    \hline
  \end{tabular}
\label{tab.atomion}
\end{table}

\begin{figure}[t]\centering
\includegraphics*[width=0.8\linewidth]{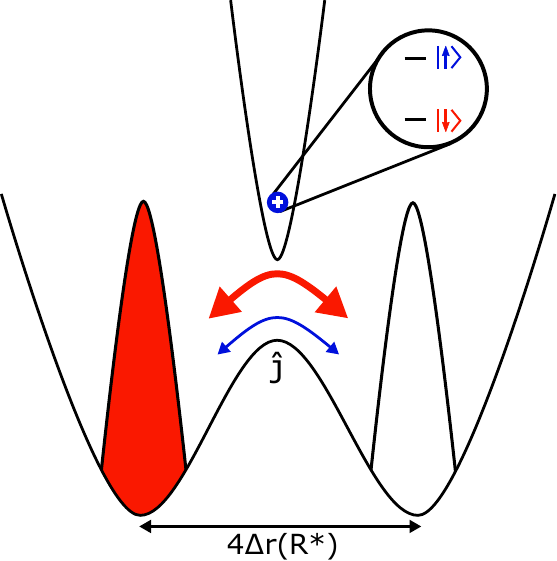}
\caption{A Bose-Einstein condensate trapped in a double-well potential with an ion in the center. The internal state of the ion is controlled by laser light and controls the tunneling rate $\hat{J}$.}
\label{fig:DoubleWell}
\end{figure}

Our setup is aiming for studies of the tunneling dynamics of a bosonic Josephson junction coupled to a single or a string of trapped ions (Fig.~\ref{fig:DoubleWell}). A single ion trapped in the center of the atomic double-well potential can control the atomic tunneling rate via spin-dependent atom-ion interactions \cite{Gerritsma.2012}. Since in this proposal a trapped ion can control many-body tunneling dynamics, mesoscopic entanglement between the atomic matter wave and the spin of the ion may be created. The interplay between the spin-dependent tunneling and the inter-atomic interactions could also result in superposition of quantum self-trapping \cite{Albiez.2005} and Josephson tunneling. Single, neutral impurities have also been proposed as a means to control and measure the dynamics in Josephson junctions, but these typically lack the sub-$\mu$m controllability offered by the trapped ion system. We can make tightly confined atomic clouds that have small sizes. This allows to limit the interactions between atoms and ions to the central region of the ion trap, where micromotion induced heating is small \cite{Holtkemeier.2016}. It may also put the 1D regime of atom-ion systems within reach \cite{Idziaszek.2007}, where the trapped ions may induce large effects in the atomic cloud, such as density bubbles \cite{Goold.2010}.

There are several reasons why mixtures of Rb and Yb were selected. i) The excessive micromotion of ion can be minimized by choosing a large ion to atom mass ratio; Yb and Rb atoms are thus a good choice for our experiment \cite{Cetina.2012}. ii) These two species are also thermodynamically favored. Yb$^{+}$ ions immersed in a Rb atomic cloud can be sympathetically cooled \cite{Zipkes.2010b}. iii) Yb has seven stable isotopes; including five bosonic and two fermionic species with nuclear spin 1/2 and 5/2 \cite{Audi.2003} which can be used as a quantum sensor for very weak forces \cite{Ivanov.2016}. iv) RbYb ground state molecules have large electric dipole moments, allowing for long range interactions \cite{Nemitz.2009}. v) It is easy to implement a mMOT and an Ioffe-Pritchard chip trap for Rb atoms; working with Rb at IR wavelengths does not affect trapped Yb$^{+}$ ions. Rb atoms will be trapped as a spin-polarized sample, thus ideal to follow this proposal \cite{Schmidt-Kaler.2012}.

\begin{figure}[t]
\centering
\includegraphics*[width=1\linewidth,angle=0]{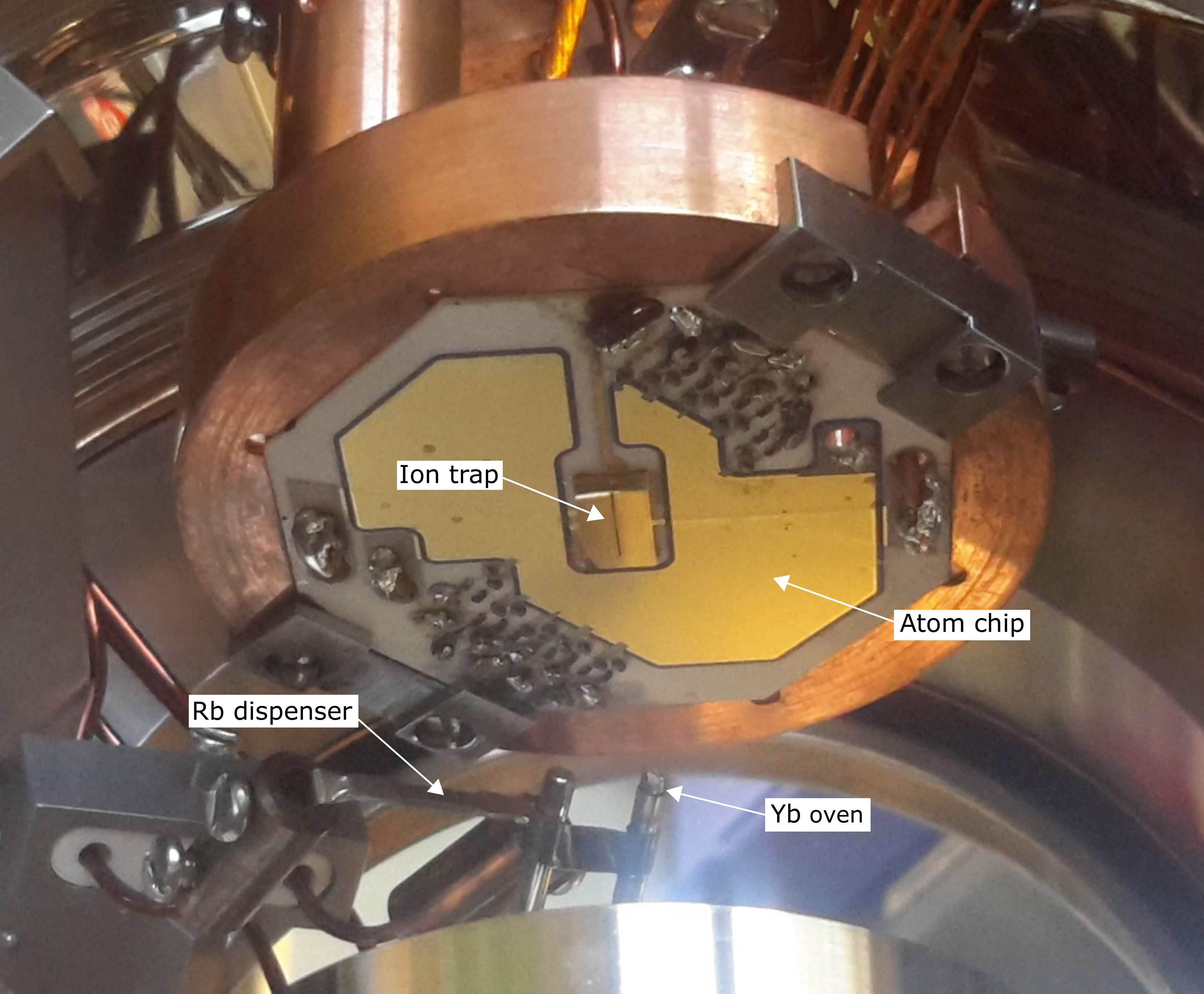}
\caption{Final assembly of the atom-ion chip with all wire-bonds. The atom source carrier is placed below the trap carrier and it includes a Rb atom dispenser and the Yb atom oven.}
\label{fig:HBtrap}
\end{figure}

The paper is structured as follows: First, we give an overview of the experimental design, then present the atom trap, the ion trap, its operation with successful trapping of Yb$^{+}$ ion crystals and discuss the characterization of the ion trap. Finally we give an outlook of our upcoming experiments. Initial experiments in the hybrid system of interacting ions and atoms have been plagued by micromotion induced heating \cite{Cetina.2012} such that we address this issue specifically in our setup.

% SYSTEM OVERVIEW %%%%%%%%%%%%%%%%%%%%%%%%%%%%%%%%%%%%%%%%%%%%%%%%%%%%%%%%%%%%%%%%%%%%%%%%
%%%%%%%%%%%%%%%%%%%%%%%%%%%%%%%%%%%%%%%%%%%%%%%%%%%%%%%%%%%%%%%%%%%%%%%%%%%%%%%%%%%
\section{System overview}
\label{sec.system}

Our setup consists of a hybrid atom-ion chip trap that can trap laser cooled neutral $^{87}$Rb atoms in a mirror-magneto optical trap (mMOT) \cite{Reichel.1999} and transfer them into a tight Ioffe-Pritchard (IP) potential. The IP trap is created by a Z-shaped wire together with a bias magnetic field and we expect trap frequencies of $(\omega_{x}, \omega_{y}, \omega_{z})=2\pi\times(1170, 1167, 84)\,$Hz from our calculations. Atoms trapped in this IP potential are spatially overlapped with Yb$^{+}$ ion crystals which are loaded into a microfabricated planar Paul trap (Fig.~\ref{fig:HBtrap}). 

\begin{figure}[t]
\centering
\includegraphics*[width=0.9\linewidth]{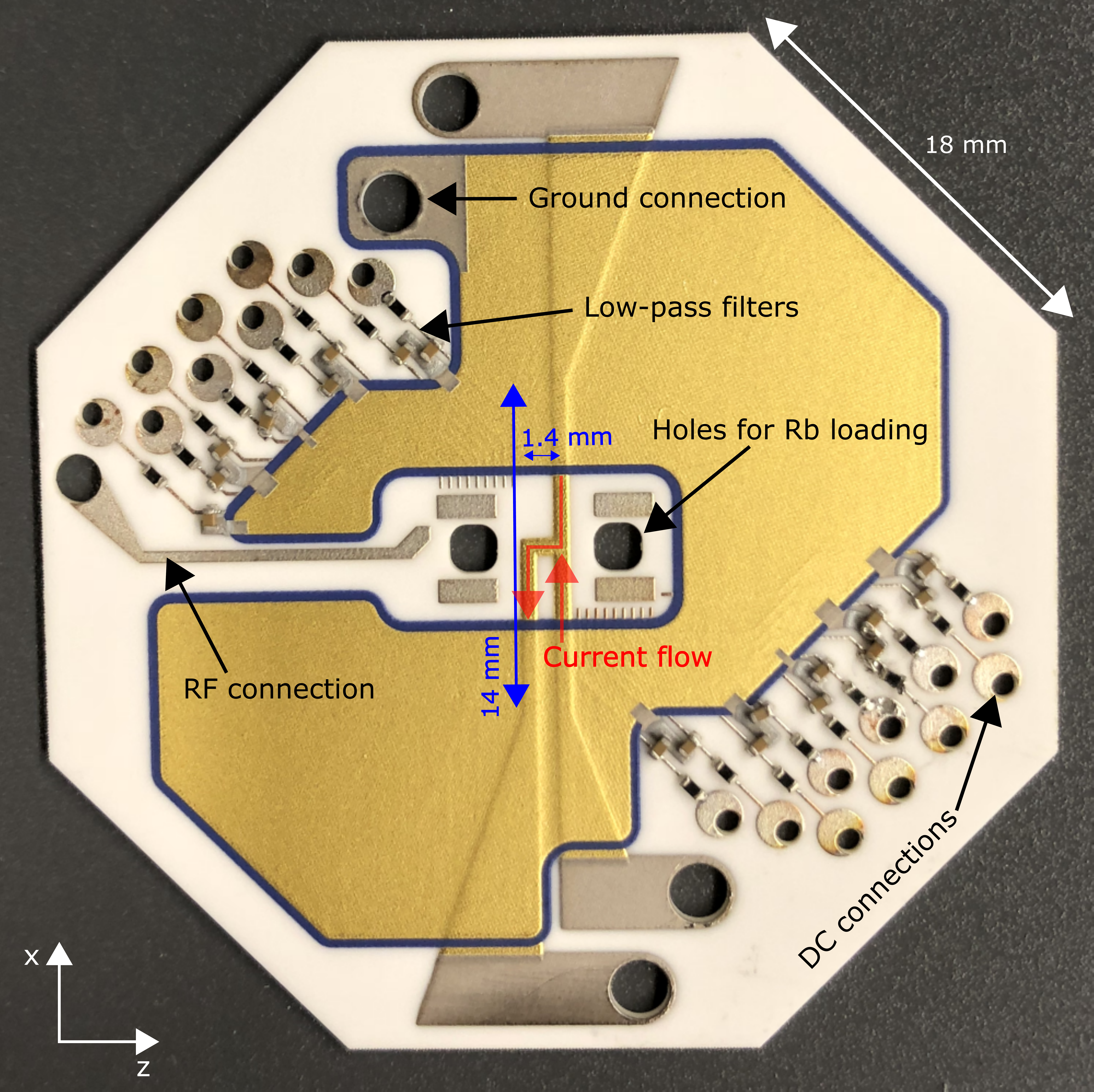} \hfill
\caption{The atom trap is composed of an octagonal chip with an outside diameter of $45\,$mm which is used as the surface for the atom trap wires, as filter board and as support of the ion trap. This direct mounting assures the required positioning accuracy. In the assembly of atom-ion chip, the ion trap is glued on top of the atom trap covering the central region of the octagon. Electrical bonding wires connect the DC and RF voltage from the octagon to the ion trap chip (Fig.~\ref{fig:HBtrap}). Underneath the octagon chip surface there are a big U-shaped, a small U- and a Z-shaped wire. It is also schematically shown how electric current can flow in the wires during the trapping phase (red arrows) to form the magnetic field configuration required for a mMOT or IP trap. The filter board includes $3.38\,$MHz low-pass filters ($C=4.7\,$nF, $R=10\,\textOmega$) on the DC control electrodes to filter RF pickups.}
\label{fig:AtomTrap1}
\end{figure}

For cooling and trapping of Rb atoms near the surface of the ion trap, we use the chip trap in a mMOT configuration \cite{Reichel.1999}. Rb atom sources are two dispensers (SAES Getters), one implemented behind and one in front of the atom-ion trap (Fig.~\ref{fig:HBtrap}). The atom trap shown in Fig.~\ref{fig:AtomTrap1} is fabricated using thick film technology in order to print UHV-compatible, sub-mm-scaled electrical circuits on the Alumina substrate (Al$_{2}$O$_{3}$). A major benefit of using this technology is the possibility of printing multiple circuit layers separated by isolating layers which we use for our design.

In order to realize a mMOT, we need two $\sigma^+$ circular polarized and two $\sigma^-$ circular polarized laser beams and a quadrupole magnetic field. Our Rb-laser system consists of two home-made external cavity diode lasers (Panasonic LNC728PS01WW) with a maximum output power of $200\,$mW at continuous wave condition. These lasers are individually frequency stabilized to the atomic transitions in the D2 line of $^{87}$Rb ($5^{2}S_{1/2}\rightarrow5^{2}P_{3/2}$) through frequency modulation spectroscopy. The cooling laser frequency is locked to the crossover transition CO(2,3) and the repump laser to the $F=1\rightarrow F'=2$ transition.  To extract information from an atomic cloud, we use an additional laser beam for absorption imaging \cite{Mewes.1996}. We take time-of-flight (TOF) images with a probe beam pulse resonant with the $5^{2}S_{1/2}(F=2)\rightarrow5^{2}P_{3/2}(F=3)$ transition, which casts a shadow on a CCD camera (Pco.pixelfly with $1392\times1040\,$px$^{2}$). To get a good signal the probe beam intensity must be below saturation which is for $\pi$-polarized light $I_{\text{sat}} = 2.50\,$mW$/$cm$^{2}$ with the resonant cross section $\sigma = 1.9\times 10^{-9}\,$cm$^{2}$ \cite{Steck.2010}.

\begin{figure*}[t]% this start makes figure wide in one coloumn
\centering
\includegraphics[width=.3\textwidth]{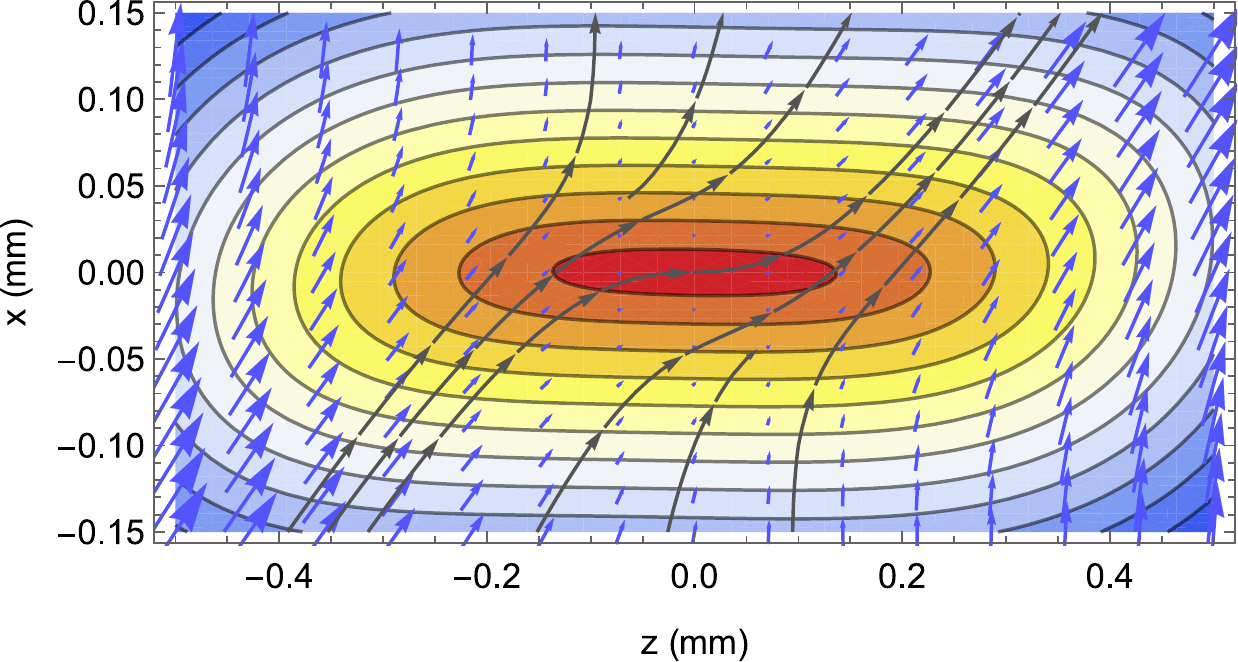} 
\includegraphics[width=.3\textwidth]{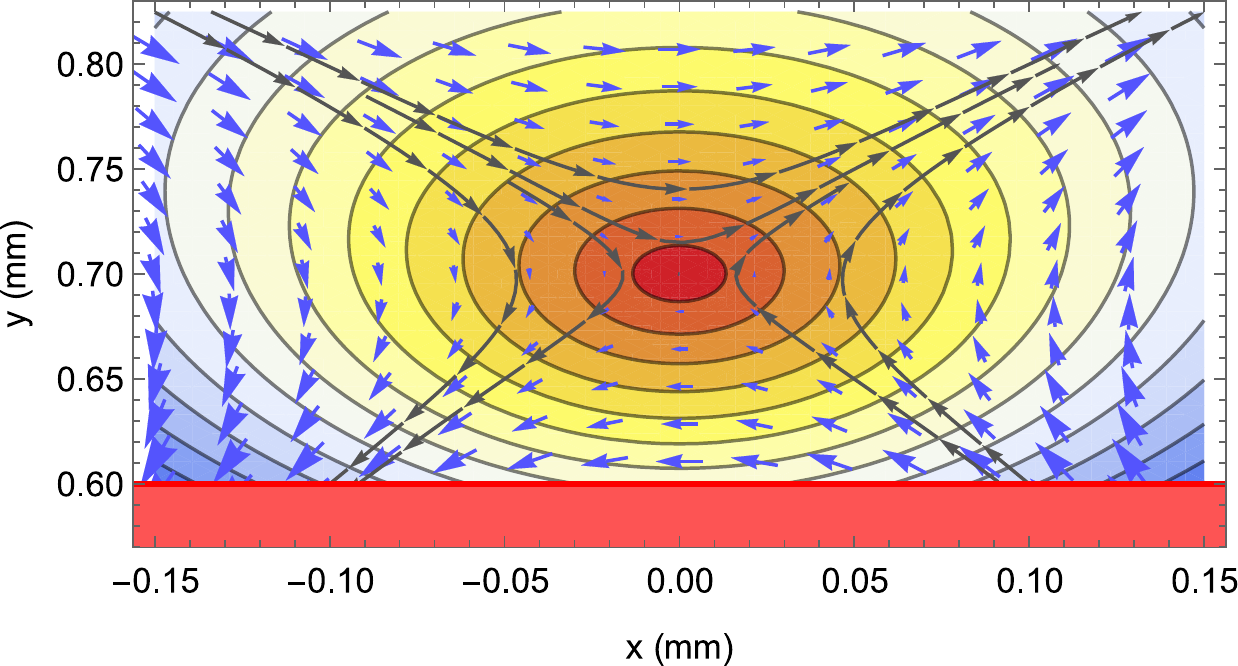} 
\includegraphics[width=.3\textwidth]{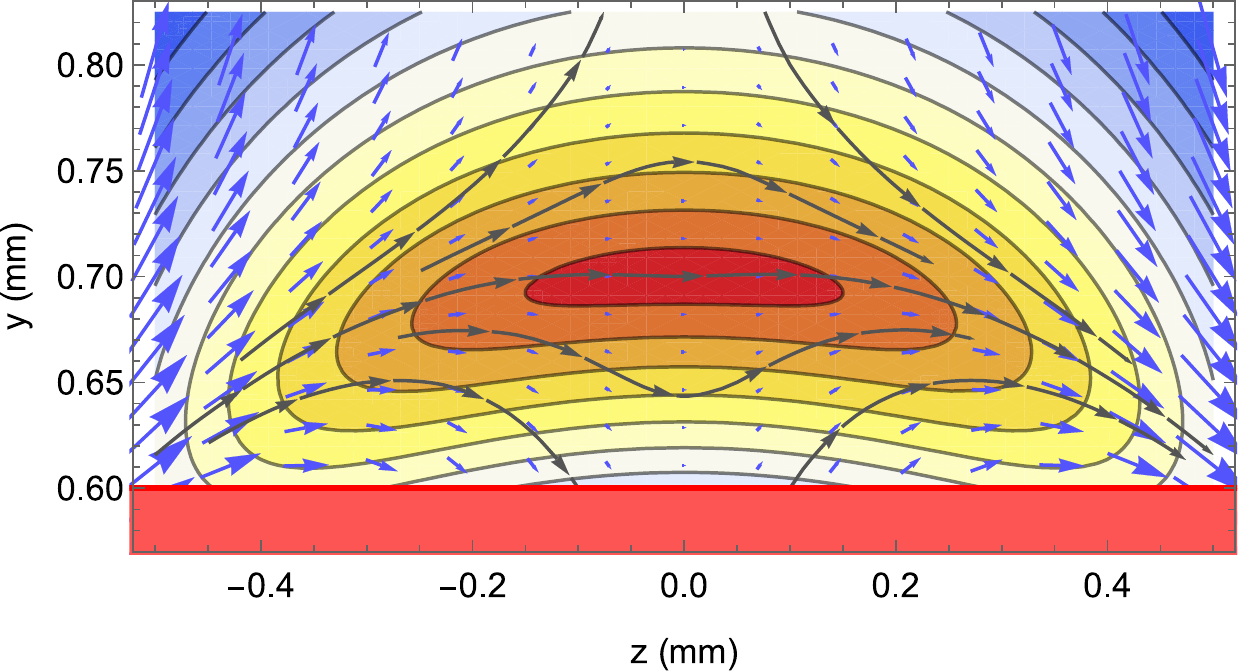}\\
\includegraphics[width=.3\textwidth,]{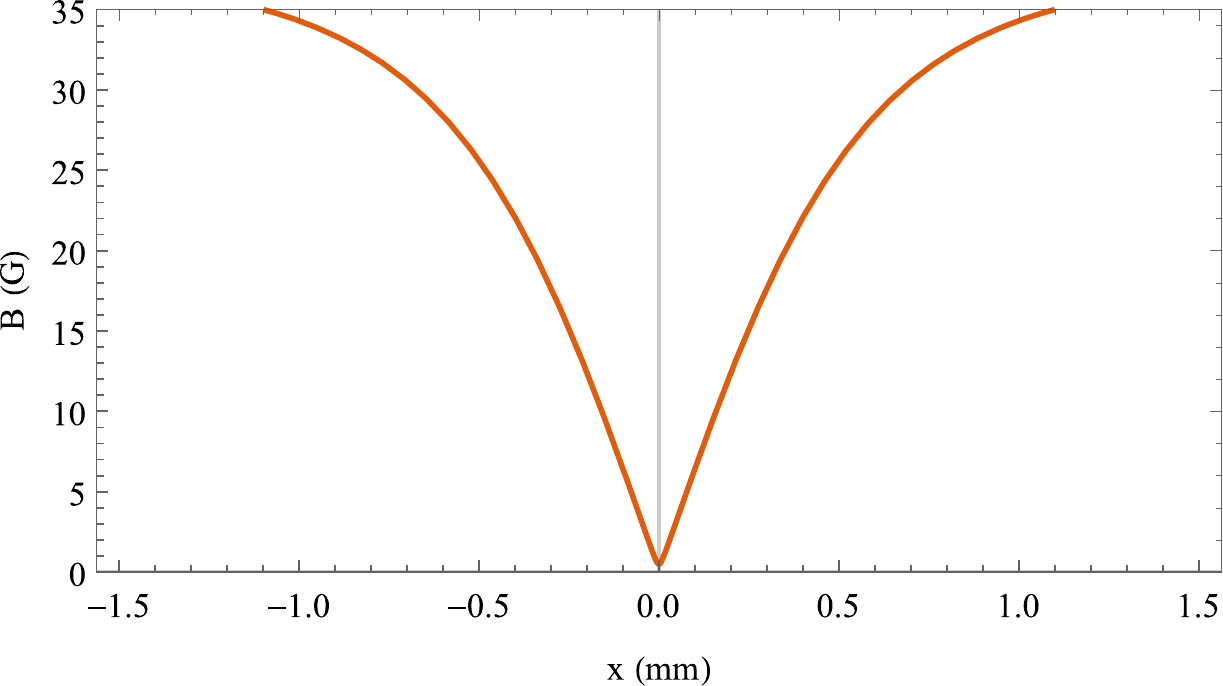} 
\includegraphics[width=.3\textwidth]{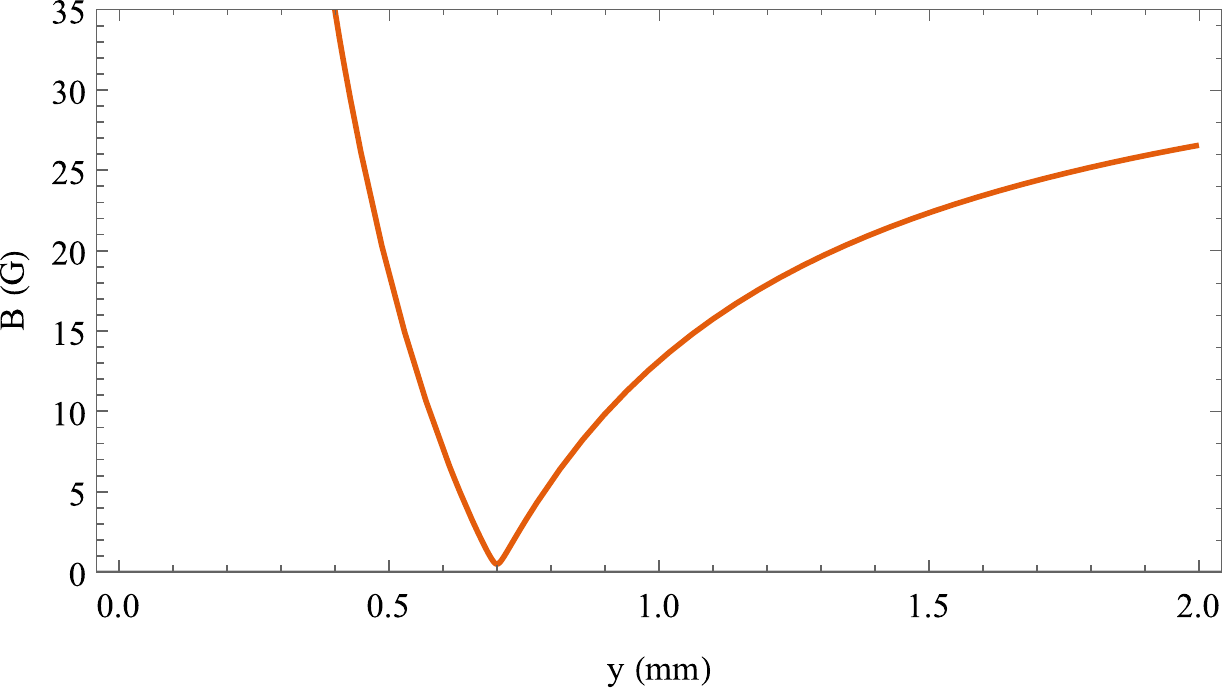} 
\includegraphics[width=.3\textwidth]{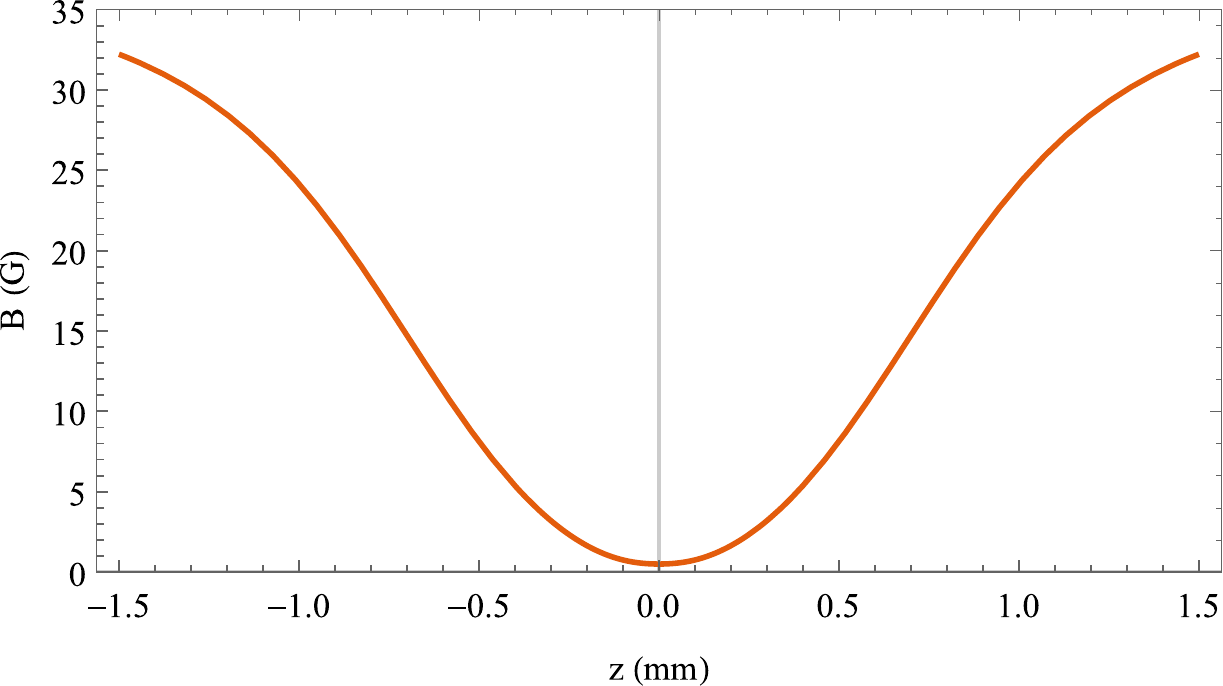}
\caption{Potentials and $B$-fields of the Z-shaped wire in z-x, x-y and z-y planes with minimum at $x_{0}=0\,$mm, $y_{0}=0.070\,$mm and $z_{0}=0\,$mm. Every color-step in the contour plots corresponds to $1$  Gauss.}
\label{fig:zwire}
\end{figure*}

\begin{figure}[t]
\centering
\subfloat[]{
\includegraphics*[width=0.45\linewidth]{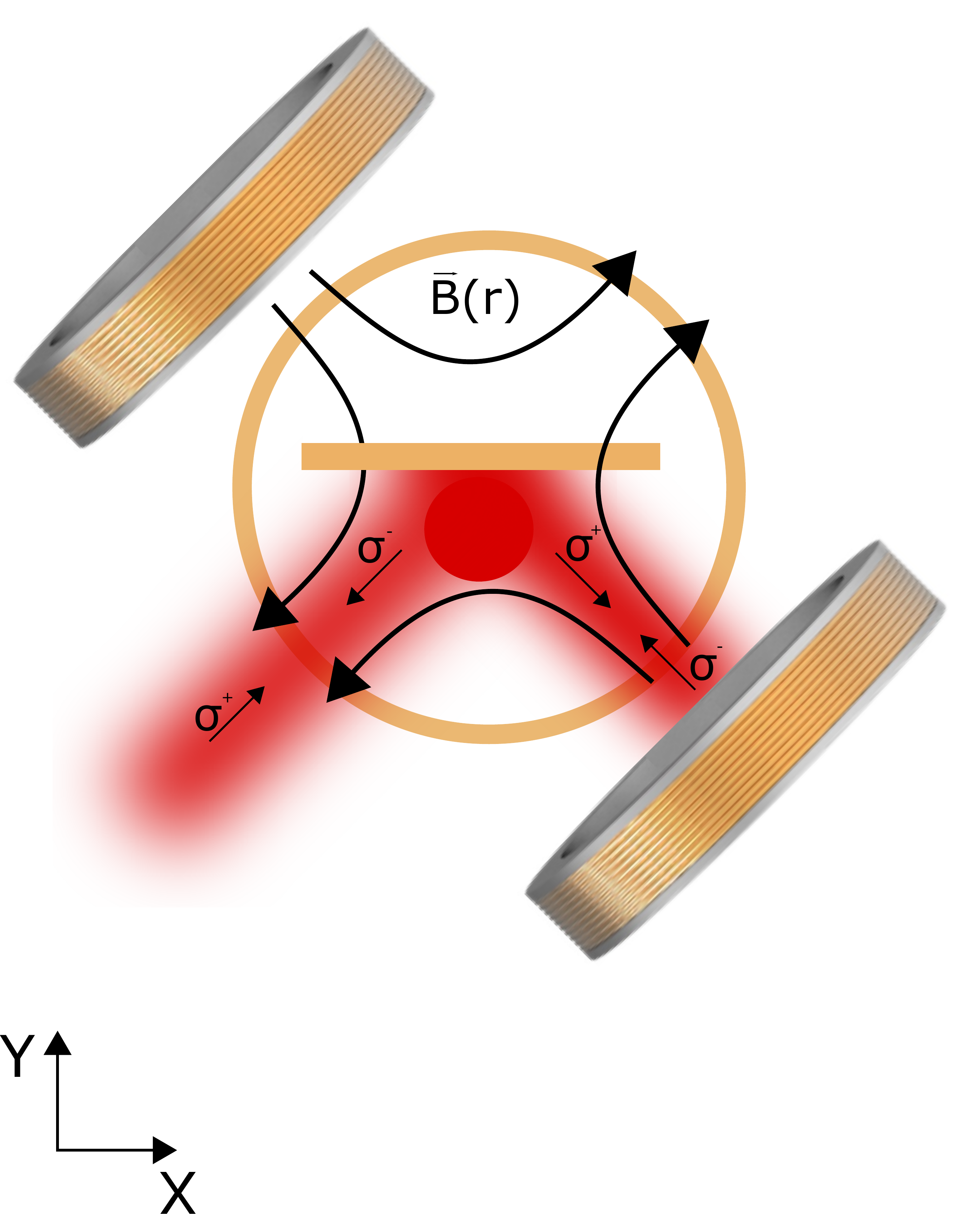}}
\subfloat[]{
\includegraphics*[width=0.5\linewidth]{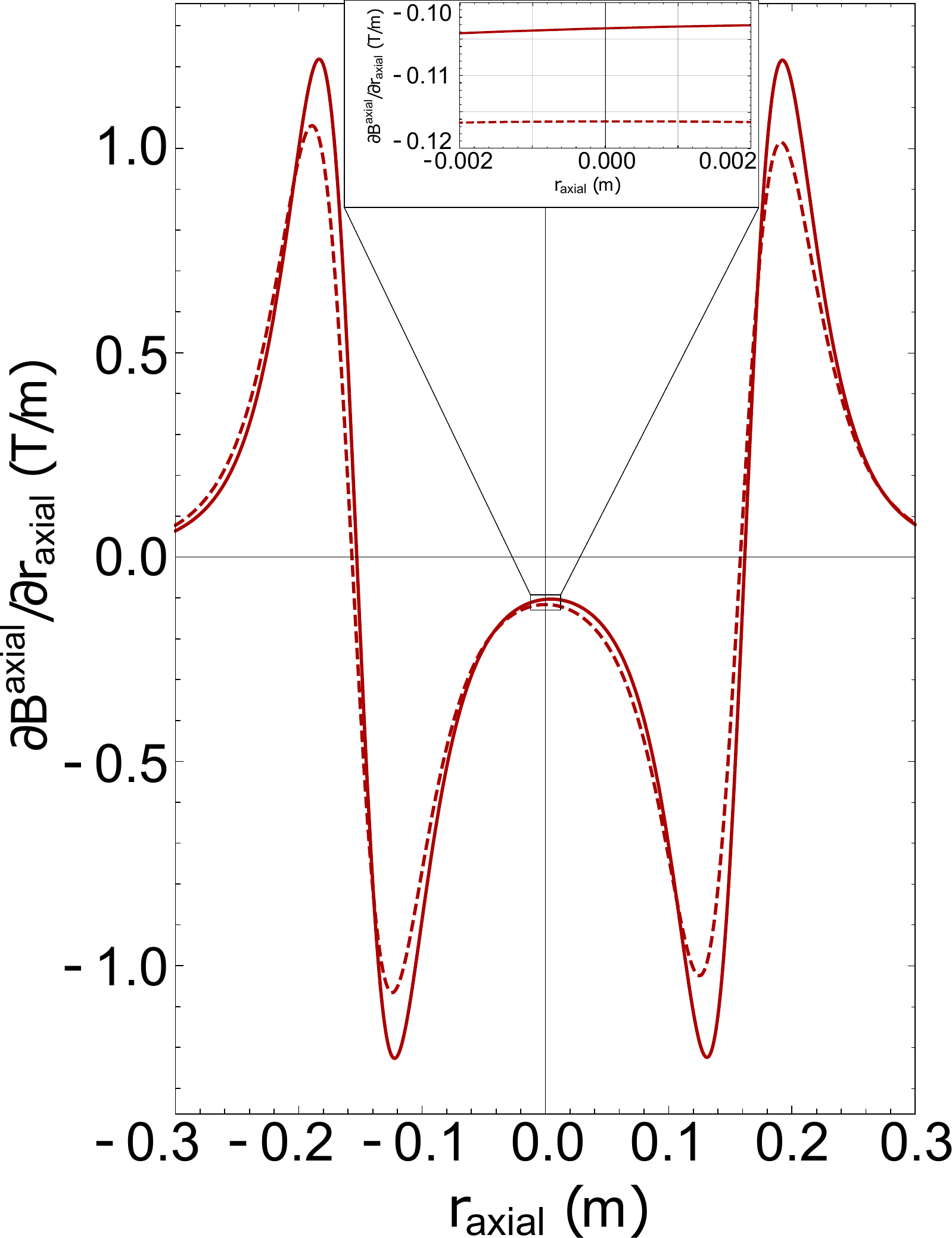}} \hfill
\caption{(a) The configuration of the laser beams and the quadrupole magnetic field to form a mMOT. (b) The magnetic field gradient along the axial direction of mMOT coils. The dashed line is the simulation using the Mathematica package Radia \cite{Radia} and the solid line is the measurement.}
\label{fig:mMOTcon}
\end{figure}

The axially symmetric quadrupole field is generated by a pair of current carrying coils (mMOT coils) in anti-Helmholtz configuration. The mMOT coils are $45^{\circ}$-rotated with respect to the plane of the trap. The distance between the two coils is $27\,$cm and each coil consists of 36 windings of a hollow-core copper wire with a quadratic cross section (outer cross section $6\times6\,$mm$^{2}$ and an inner cross section $4\times4\,$mm$^{2}$) wrapped in Kapton tape for electrical insulation. These coils are internally cooled down to 20$^{\circ}$C with a $300\,$Watts chiller (MINORE 0-RB400). At a maximum current of $200\,$A supplied by a high power source (SM 30-200 Delta Elektronica), we measure a magnetic field gradient of $\partial B^{\text{axial}}/\partial r_{\text{axial}}=0.11\,$T$/$m at the center of the trap which is at a few mm from the ion trap surface (Fig.~\ref{fig:mMOTcon}). With a one winding coil behind the holder we can adjust the field, to push or pull the atom cloud in respect to the trap surface. For propper trapping efficiency measured magnetic field gradients are in the range of $0.10-0.20\thinspace$T/m \cite{RbLoad2001}.

Once atoms are captured in the mMOT, the mMOT coils will be switched to a bias field while switching on the big U-shaped wire beneath the atom chip which generates a quadrupole field with a minimum at infinity. The atoms can now be confined at about $2\thinspace$mm underneath the chip surface in a spatially smaller area. The atoms will then be transferred to a potential created by the small U-shaped wire in the atom chip. At this step we shift the atoms close to the ion trap surface and compress the atom cloud to a smaller and steeper mMOT volume. As the final step we create an Ioffe-Pritchard trap by turning on the Z-shaped wire in the atom chip together with a bias magnetic field (Fig.~\ref{fig:zwire}). The axial direction of the atom cloud trapped in the magnetic field of the Z-shaped wire is not necessarily parallel to the z-axis of the ion trap. In the atom trap design we have considered the length of the Z-shaped wire to be $1.4\,$mm, so that the atom cloud overlaps with the ion cloud.

\begin{figure}[t]
\centering
\includegraphics*[width=1\linewidth,angle=0]{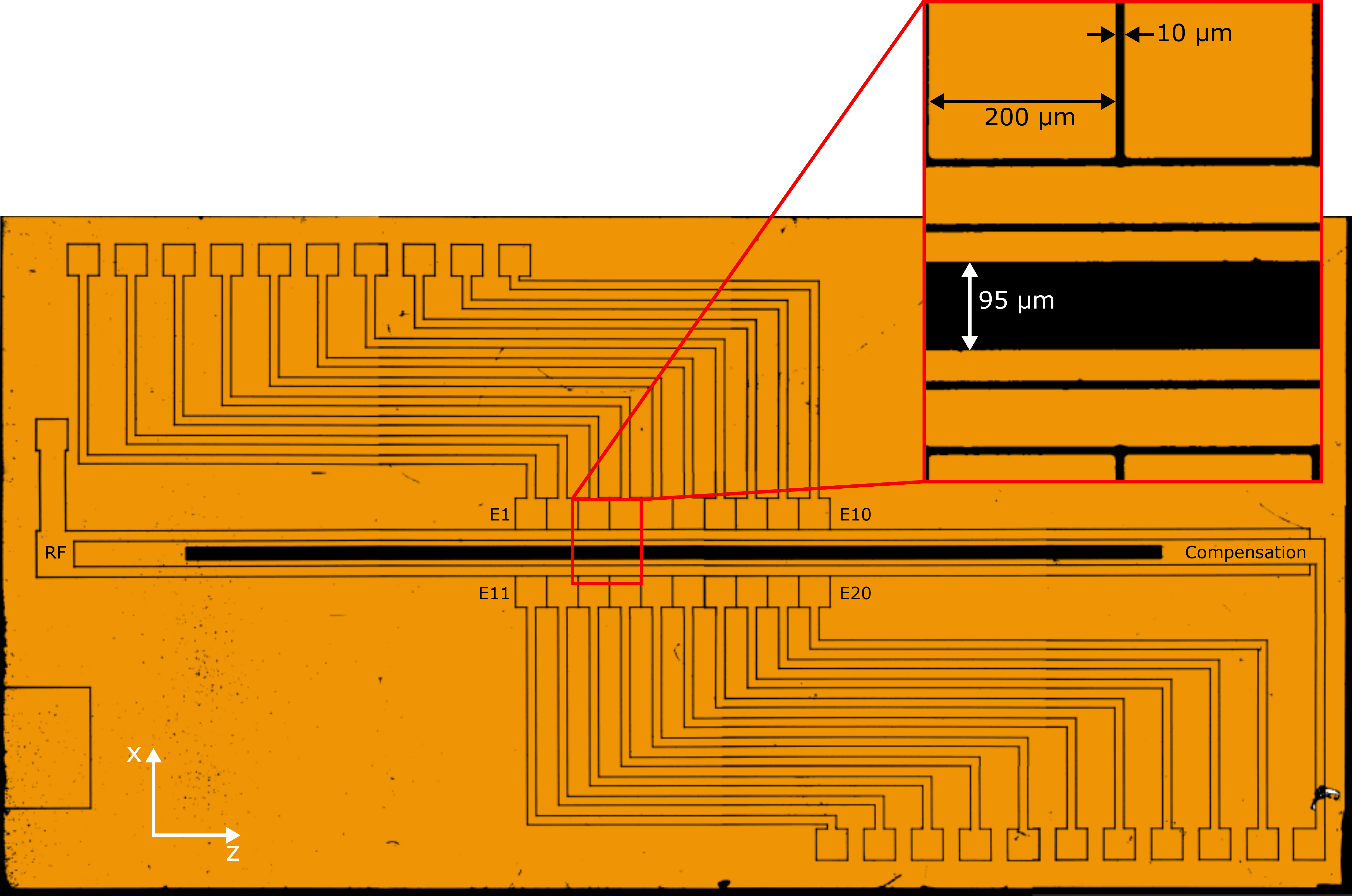}\\
\caption{Optical microscope image of the microfabricated surface trap used in our experiment. The chip has a size of $9\times 4.5\,$mm$^{2}$ and a thickness of $500\,\mu$m. It has a long and symmetric F-shaped rail for RF confinement, 20 static-voltage electrodes with a size of $200\times 200\,\mu$m$^{2}$ used for the radial confinement (E01 - E20) and one inner compensation electrode which is axially and symmetrically extending along a slit of $100\,\mu$m width and $5\,$mm length (E21). This slit is used to load Rb atoms from a dispenser placed right behind the ion trap. In the experiments described here, Yb$^{+}$ ion crystals are trapped and confined along the trap axis (z-direction). The gaps between the electrodes are approximately $10\,\mu$m wide and $50\,\mu$m deep, which are large enough that the RF voltage does not cause electrical breakdown at $100-200\,$V$_{\text{pp}}$.}
\label{fig:iontrap}
\end{figure}

\begin{figure}[t]
\centering
\includegraphics[width=1\linewidth]{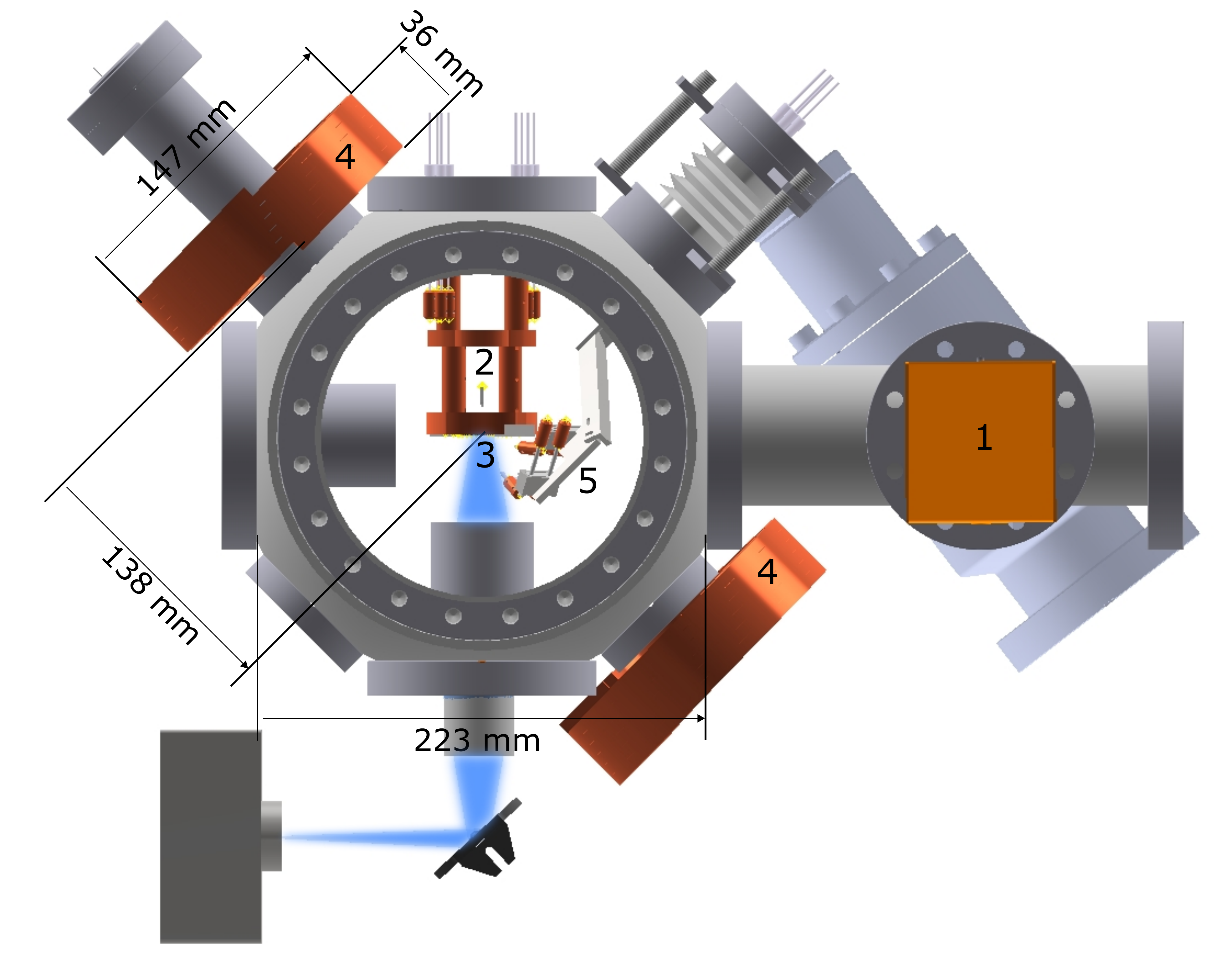}%Carefull with this ref, vacuum is wrong. I don't change it, to not destroy crossrefs.
 \caption{Overview of the system: 1. NEG pump, 2. Trap carrier with Rb atom source, 3. Hybrid trap, 4. mMOT coils, 5. Rb and Yb atom sources. The trap is mounted upside down on a home-made carrier which is attached to a special CF63 flange featuring various electrical feedthroughs. The trap surface is right in the middle of the vacuum chamber. Using a weld bellow (CF35/SEWB) one can adjust the atom sources relative to the trap chip.}
\label{fig:vaccum}
\end{figure}

The ion trap we use in the experiment (Fig.~\ref{fig:iontrap}) is a microfabricated surface trap with 21 DC electrodes which is fabricated in Translume technology \cite{Daniilidis.2014,Amadeo.2015}. A pulsed laser is used to pattern trenches of $50\,\mu$m deep and $10\,\mu$m wide in a fused silica substrate (SiO$_{2}$). Then material is etched with hydrofluoric acid which has a strong corrosive characteristic against SiO$_{2}$. Finally the surface is coated with four metal layers of $20\,$nm Titanium, $150\,$nm gold, $20\,$nm Titanium and $150\,$nm gold, i.e. $300\,$nm gold in total. The high quality gold coating has a reflectance of about 90\% for 780~nm light at an incident angle of 45$^{\circ}$. According to successful operations of mMOTs on a microfabricated atom chip \cite{Folman.2000}, and also on a magnetic lattice with periodicity of $10\,\mu$m \cite{Hannaford.2014}, we expect a similar performance in our system.

The ion trap is glued on top of the atom trap using a UV adhesive (EPO-TEK$^{\textregistered}$ OG142-112 UV Cure Optical Epoxy). The adhesive is carefully applied at the edges of the ion trap while preventing the adhesive to run between the two traps and is cured by illuminating it three times for $10\,$s with UV light. The electric connections between the ion trap and the filterboard are established by wire bonds ($25\,\mu$m diameter Pd wire).  Planar Paul traps with characteristic length of about 100$\thinspace\mu$m have advantages in terms of fabrication and scalability \cite{Kielpinski.2002,Madsen.2004,Seidelin.2006,Pearson.2006,Leibrandt.2007,Wang.2009,Amini.2011}. In planar traps the ions are confined radially by RF potentials and axially by static electric voltages (DC) and all electrodes lie in a single plane \cite{Chiaverini.2005}. There are several analytical approaches like a model by House \cite{House.2008} and Biot-Savart-like methods \cite{Oliveira.2001,Wesenberg.2008,Schmied.2010} to design planar traps with arbitrary geometries. In this report we use the House model for analytical determination of the applied DC and RF potentials to trap Yb$^{+}$ ions. The assembled trap is shown in Fig.~\ref{fig:HBtrap} where the atom-ion chip is mounted upside-down on a carrier which is attached to a CF63 having various electrical feedthroughs. The chip surface is right in the middle of the vacuum chamber which guarantees optimal optical access \cite{Joger.2013}. Fig.~\ref{fig:vaccum} shows the complete vacuum setup and its main functional parts. Seven anti-reflection coated fused quartz silica viewports allow laser access to the center of the ion trap and the atom trap. The distance between the hybrid trap and the surface of the big window is $85\,$mm. To reach a UHV pressure, an ion getter pump (NEXTorr$^{\textregistered}$D200-5) is attached to a cross right next to the chamber. The pressure we reach in our chamber is $1.3\times10^{-9}\,$hPa.
% PLANAR ION TRAP %%%%%%%%%%%%%%%%%%%%%%%%%%%%%%%%%%%%%%%%%%%%%%%%%%%%%%%%%%%%%%%%%%%%%%%%
%%%%%%%%%%%%%%%%%%%%%%%%%%%%%%%%%%%%%%%%%%%%%%%%%%%%%%%%%%%%%%%%%%%%%%%%%%%%%%%%%%

\section{Observation of ions}
\label{sec.iontrap}

We introduce our surface trap which has a F-shaped RF electrode for axial confinement of the ions. For the radial confinement it has 20 DC electrodes, which are placed axially symmetric around the RF rail and one compensation electrode in-between the legs of the RF. The optical resolution of our imaging system is about $720\,$nm and we typically trap Yb$^{+}$ ions at a distance of $150\,\mu$m to the surface of the trap and can position them axially with an accuracy of about $\pm\,80\,$nm, see Sec. \ref{sec:TrapOp}.

% TRAP OPERATION%%%%%%%%%%%%%%%%%%%%%%%%%%%%%%%%%%%%%%%%%%%%%%%%%%%%%%%%%%%%%%%%%%%%%%%%%
%%%%%%%%%%%%%%%%%%%%%%%%%%%%%%%%%%%%%%%%%%%%%%%%%%%%%%%%%%%%%%%%%%%%%%%%%%%%%%%%%%
\section{Ion trap operation}
\label{sec:TrapOp}
The configuration of our Paul trap is linear, thus radial confinement is achieved by applying a RF sinusoidal drive field. Ions are confined in the node line of the quadrupole field that is generated from the electrodes. This harmonic binding in the radial r$_1$ and r$_2$ directions is described by a pseudopotential $\textPhi_{\text{RF}}$. Additionally, ions or ion crystals, are confined along the trap z axis by a DC harmonic oscillator potential $\textPhi_{\text{DC}}$. In the full description, both, RF and DC fields are independently adjusted for controlling the ion crystals' position and alignment, the inter-ion distances, and the trapping frequencies in all directions. The challenge is to find control voltages that suite the experimental protocol, and also match the condition that the center of the DC potential falls on the RF node line, where the ion micromotion is compensated (see Sec. \ref{subsec:micro}).

The total effective potential for trapping the ions $\textPhi\left(\boldsymbol{r},\thinspace t\right)$ is given by the sum of a time-independent potential $\textPhi_{DC}$ generated by the trap DC electrodes and a sinusoidally varying part, the pesudopotential $\textPhi_{RF}$, that is driven by a RF voltage source:

\begin{eqnarray}
	\textPhi\left(\boldsymbol{r},\thinspace t\right)	&=& \textPhi_{RF}+\textPhi_{DC} \\
	&=& \frac{q^{2}V_{RF}^{2}}{4M\textOmega_{RF}^{2}}\left|\left|\nabla\phi_{RF}\right|\right|^{2}+Q\sum V_{i}\phi_{DC,\thinspace i},
\end{eqnarray}

here, $q$ and $M$ are the charge and mass of the trapped ions. $V_{RF}$ and $\varOmega_{RF}$ denote the RF amplitude and frequency, respectively. The potential terms $\phi_{RF}$ and $\phi_{DC}$ are the solution of the Laplace equation for a unit voltage applied to the RF and the $i$th DC electrode. V$_{i}$ is the voltage applied on the $i$th DC electrode. The radial confinement of ions is realized by applying a RF voltage at a frequency of $\textOmega_{RF}=2\pi \times11.22\,$MHz and an amplitude of $109\,$V$_{\text{pp}}$. This RF signal is generated by a RF generator (Rohde $\&$ Schwarz SML01), amplified (ZHL-5W-1-Mini Circuits) and sent to a helical resonator with a Q-factor of $360$ to further enhance the signal \cite{Siverns.2012}. With these values we reach a relatively small trap depth of $50\,$meV. In order to filter RF pickups, we have implemented a $3.38\,$MHz low-pass filter on the trap chip connected to each DC electrode and a second $50\,$kHz low-pass filter connected to the feedthrough. At our typical RF voltages we measured RF pickups of $116-930\,$mV$_{pp}$ in the DC electrodes. Trapped ions are confined with typical trap frequencies of ($\omega_{r1},\omega_{r2},\omega_{a})=2\pi\times(485, 724, 289)$ kHz, where $\omega_{a}$ is the axial and $\omega_{r1}$ and $\omega_{r2}$ are the radial trap frequencies of the ion tap.

To provide the voltage for the DC electrodes, we use FPGA-controlled digital-to-analog conversion (DAC) boards. They allow sweeping voltages in real time and avoid delays caused by PCs. The ion transport is achieved by shifting the DC potential minimum. Therefore, ultra-fast voltage update rates need to be incorporated. This makes the use of FPGA controlled DACs necessary \cite{Walter.2012,Ruster.2014}. Hence, the design we use is mainly comprised of a high-speed and low noise advance multichannel arbitrary waveform generator (N-MCWG) developed at the University of Mainz \cite{SchmidtKaler.}. It supports up to 48 independent analog channels ($16\,$bits) and 25 digital TTL signals per unit. Analog channels have a voltage range of $-10\,$V to $+10\,$V with a resolution of ($0.12\,\text{mV}\, 1\,\text{LSB}$).

\begin{figure}[t]
\centering
\includegraphics*[width=0.9\linewidth,angle=0]{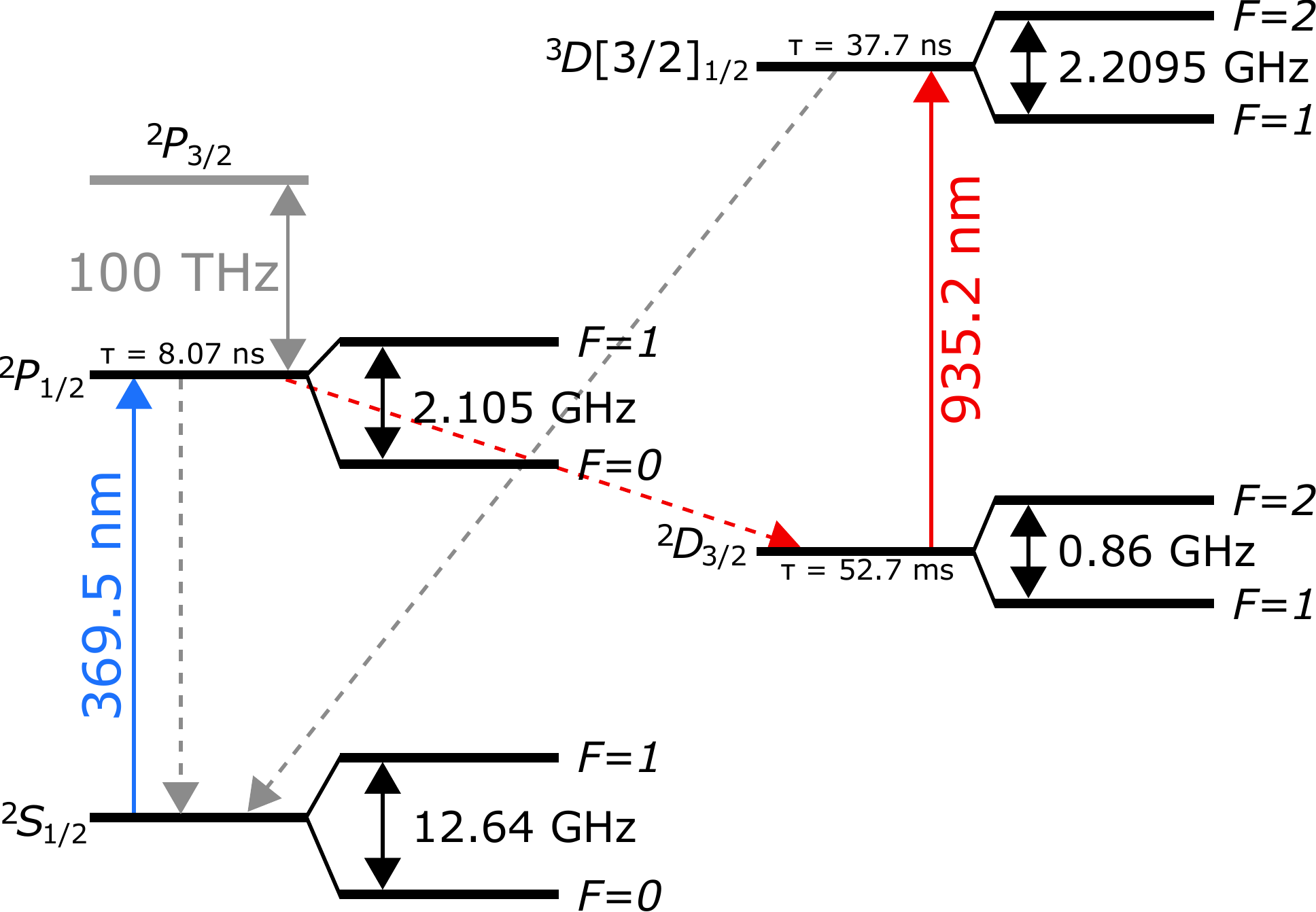}
\caption{Energy level scheme of the electronic ground and excited states of $^{171}$Yb$^{+}$ with nuclear spin I=$1/2$; the isotopes we have trapped are without nuclear spin. The transitions driven by diode lasers in our experiment are marked as straight lines. Life times and decay branching ratios are taken from \cite{Olmschenk.2009}. Additional low-lying $^{2}$D$_{5/2}$ and $^{2}$F$_{7/2}$ states have been omitted for clarity.}
\label{fig:Yblevel}
\end{figure}

\begin{figure}[b]
\centering
\includegraphics*[width=\linewidth]{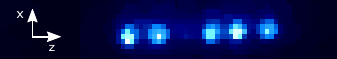}
\caption{Linear crystal of $^{174}$Yb$^{+}$ ions trapped with corresponding trap frequencies of  ($\omega_{x},\omega_{a})=2\pi\times(406, 110)$ kHz. The dark ion observed is in most of the cases $^{172}$Yb$^{+}$. Each pixel in the image corresponds to $1.09(7)\,\mu$m.}
\label{fig:5ions}
\end{figure}

The \textbf{Yb laser system} consists of three lasers used to ionize, cool and excite dipole transition of the Yb$^{+}$ ions. For the initial steps we rely on $^{174}$Yb$^{+}$, while in the further run $^{171}$Yb$^{+}$ with a long lived spin 1/2 system of ground states $^2S_{1/2}(F=0)$ and $^2S_{1/2}(F=1)$ will serve as qubit. The relevant level structure of $^{171}$Yb$^{+}$ is shown in Fig.~\ref{fig:Yblevel}. To isotope-selectively ionize neutral Yb atoms, we use a two-step photoionization scheme \cite{Aymar.1980}. First we excite the neutral Yb atoms from $^{1}$S$_{0}\rightarrow {}^{1}P_{0}$ with a home-made grating-stabilized diode laser at 398.9$\thinspace$nm (Nichia NDV4B16). The first step excitation is impinging the atomic beam under right angle and allows for isotope-selective excitation and trap loading \cite{Gulde2001,Johanning2011}. We use a Toptica TA-SHG pro as a second step to the continuum and for cooling of $^{174}$Yb$^{+}$ ions, by exciting the $^{2}$S$_{1/2}\rightarrow{}^{2}P_{1/2}$ transition ($\Gamma=2 \pi \times 19.7\thinspace$MHz), which yields to laser-induced florescence for single ion detection at a wavelength of 369.5$\thinspace$nm. Ions in the $^{2}P_{1/2}$ state can decay to the lower lying $^{2}D_{3/2}$ state with a branching ratio of $\alpha= 0.005 01(15)$~\cite{Olmschenk.2007}. Therefore, to keep the ions in the cooling cycle, we use an additional Toptica DL pro Laser at $935.2\,$nm to repump from the $^{2}D_{3/2} \rightarrow {}^{3}D[3/2]_{1/2}$ state ($\Gamma=2 \pi \times 4.2\thinspace$MHz). Both, the $369.5\,$nm and the $935.2\,$nm lasers are locked to Fabry-P{\'e}rot cavities with finesses of $630$. The $369.5\,$nm cooling and the $398.9\,$nm photoionization lasers are coupled to the same polarization maintaining fiber and the outcome beam is overlapped with the $935\,$nm repump beam using a dichroic mirror. All beams pass through a $f = 125\,$mm achromatic lens and are aligned in parallel to the surface of the trap. The beams have a slight angle of $7^{\circ}$ to the trap axis. The cooling beam has an effective beam diameter of $30\,\mu$m at its focal point and a saturation power of about $60\,\mu$W. A magnetic field of $0.13\,$mT at  $135^{\circ}$ to the laser k-vector, sets the direction of the quantization axis. The fluorescence of the ions is collected with a microscope objective with a numerical aperture of $0.35$ through an inverted viewport and detected by an EMCCD (Andor Luca $8\times8\,{\mu\text{m}}^{2}$) equipped with a bandpass filter (FB370-10); Our imaging system has a calibration factor of $1.09(7)\,\mu\text{m}$/px (Fig.~\ref{fig:5ions}).

The dark ion in Fig.~\ref{fig:5ions} is in most of the times  $^{172}\text{Yb}^{+}$ isotope. This isotope is the second most abundant isotope of  $\text{Yb}$ atom which has a photoionization energy close to the $^{174}\text{Yb}$ isotope. By trapping Yb$^{+}$ ions, we could find the cooling and repumping laser wavelengths for $^{170}\text{Yb}^{+}$, $^{172}\text{Yb}^{+}$, $^{174}\text{Yb}^{+}$ and $^{176}\text{Yb}^{+}$  which are all nuclear-spin free isotopes. The first step ionization wavelengths are extracted through the fluorescence signal of neutral Yb atoms (Tab.~\ref{tab:YbWl}). These wavelengths are in good agreement with the values stated in Ref. \cite{McLoughlin.2011}.

\begin{table}[t]
\caption{Yb transition wavelengths (vacuum). These values are measured using a HighFinesse wavelength meter (WS6). The $^2S_{1/2} \rightarrow {}^2P_{1/2}$ and $^2D_{3/2} \rightarrow {}^3D[3/2]_{1/2}$ transition wavelengths were obtained by observing fluorescence signal levels from trapped ions.}
\centering
\begin{tabular}{ccccc}
\hline
Isotope 	& Ionization (nm) & Cooling (nm) & Repump (nm) 	\\\hline
$^{170}$Yb  & $398.91070(5)$	 & $369.52360(5)$ & $935.19750(5)$\\
$^{172}$Yb	 & $398.91100(5)$	&$369.52440(5)$	&$935.18730(5)$\\
$^{174}$Yb  & $398.91135(5)$	 & $369.52495(5)$	& $935.18015(5)$\\
$^{176}$Yb	& $398.91160(5)$	&$369.52550(5)$	&$935.17240(5)$\\\hline
\end{tabular}
\label{tab:YbWl}
\end{table}

We measured the lifetime of a single ion under continuous Doppler cooling and in the dark  (Tab.~\ref{tab:lifetime}). The lifetime of a Doppler cooled single ion is usually many hours but in our experiment we measured short lifetimes of $2\,$min only due to relatively high background gas pressure of $1.3\times10^{-9}\,$hPa. Nevertheless these values are in good agreement with other shallow surface Paul traps at room temperature \cite{Leibrandt.2009}. The lifetime of a single ion in dark is about $1\,$min which is long enough for atom-ion interaction studies. In typical sequences, modifications of the loss rate are an observable at time scales of about $100\,\text{ms}$ \cite{Joger.2017}. 

\begin{table}[b]
\caption{$1/e$ lifetimes ($\tau$) of a single $^{174}$Yb$^+$ ion trapped in different axial positions of the trap.}
\centering
\begin{tabular}{p{.3\linewidth}p{.3\linewidth}p{.3\linewidth}}
\hline
Ion position	&$\tau$ with cooling (s)        	&	 $\tau$ w/o cooling (s)	\\ \hline
E02, E12	&$198.4 \pm 0.5$				&	$50.5 \pm 4.3$    \\
E04, E14	&$128.6 \pm 0.8$				&	$40.6 \pm 3.3$\\
E06, E16	&$89.5 \pm 0.5$			    &	$39.9 \pm 2.7$\\
E08, E18	&$151.6 \pm 0.6$				&	$31.7 \pm 2.6$\\ \hline
\end{tabular}
\label{tab:lifetime}
\end{table}

\begin{figure}[t]
\includegraphics*[width=\linewidth]{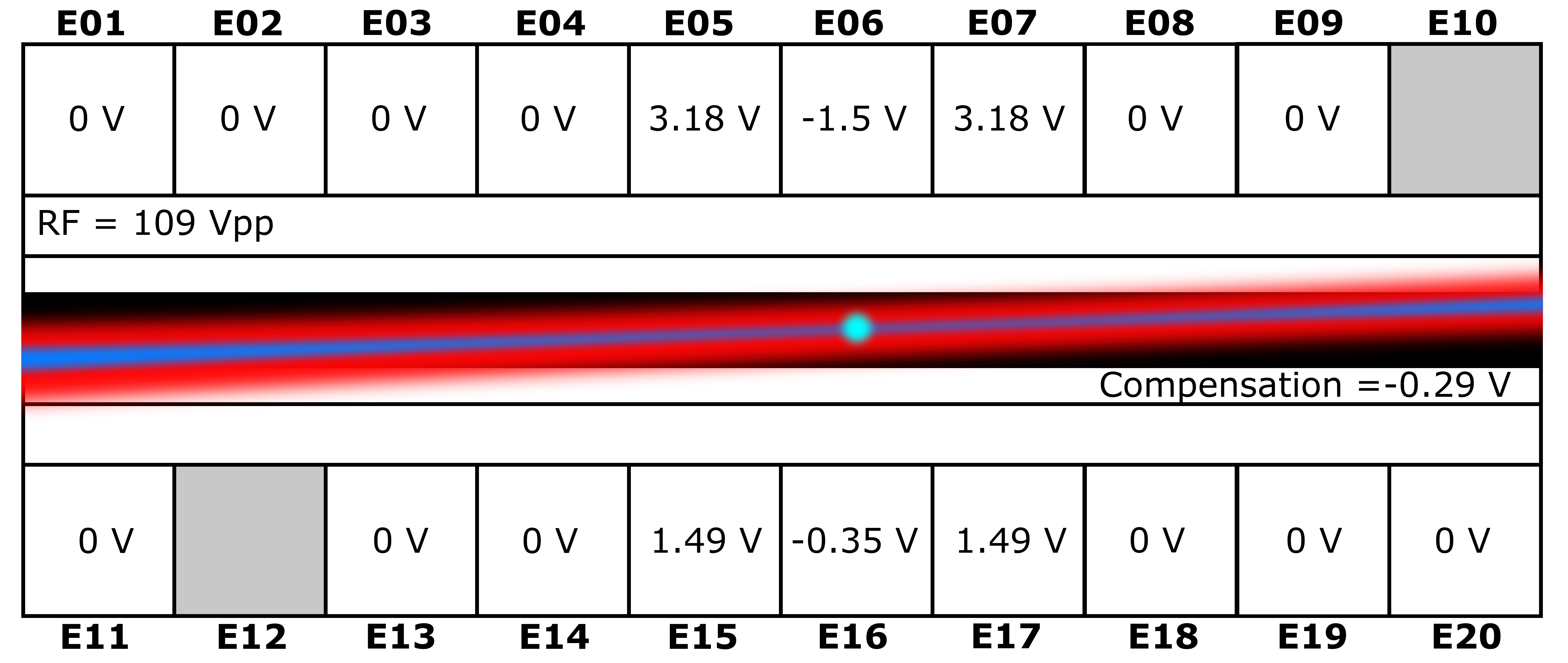}
\caption{Simulated voltages to produce a tilted potential between electrodes E06 and E16 where the Yb$^+$ is trapped. The alignment of Gaussian beams are also schematically shown here.}
\label{tab:SimUnH}
\end{figure}

\begin{figure}[t]
\centering
\subfloat[]{
\includegraphics*[width=0.5\linewidth]{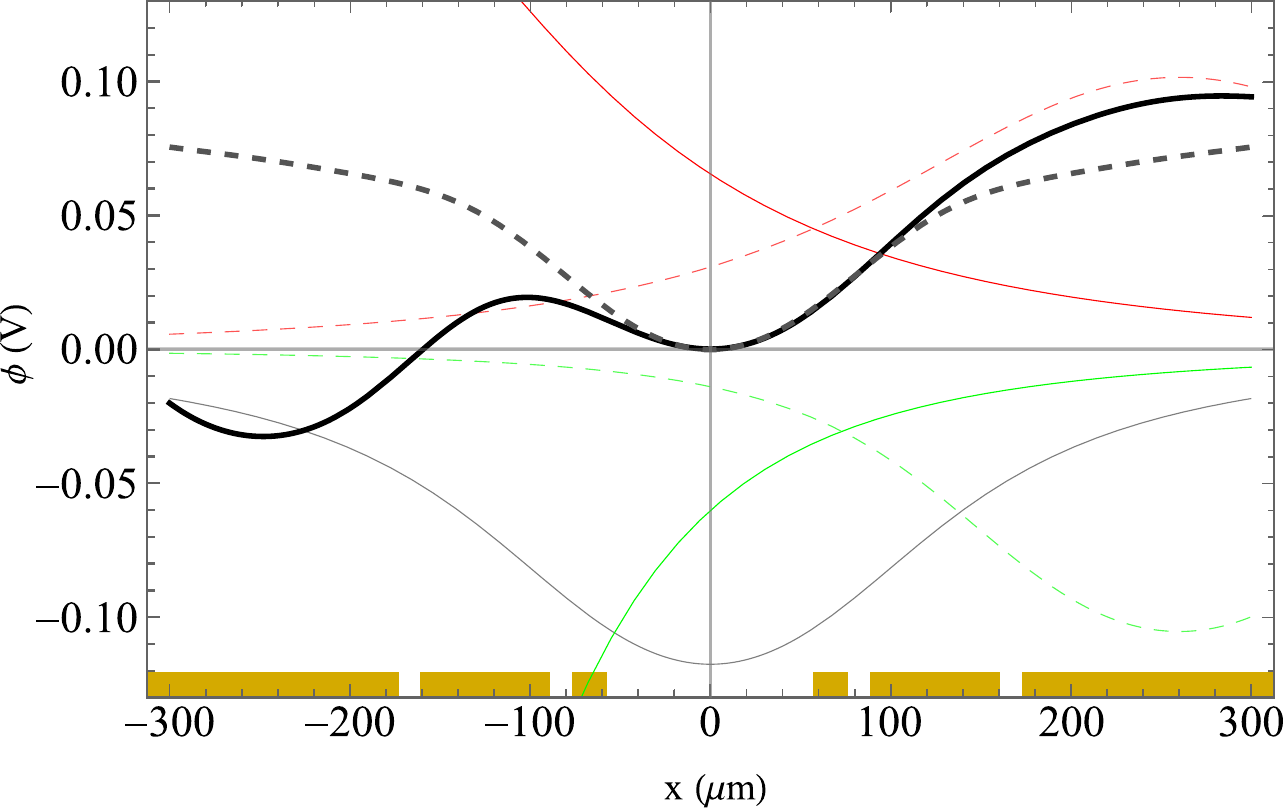}}
\subfloat[]{
\includegraphics*[width=0.5\linewidth]{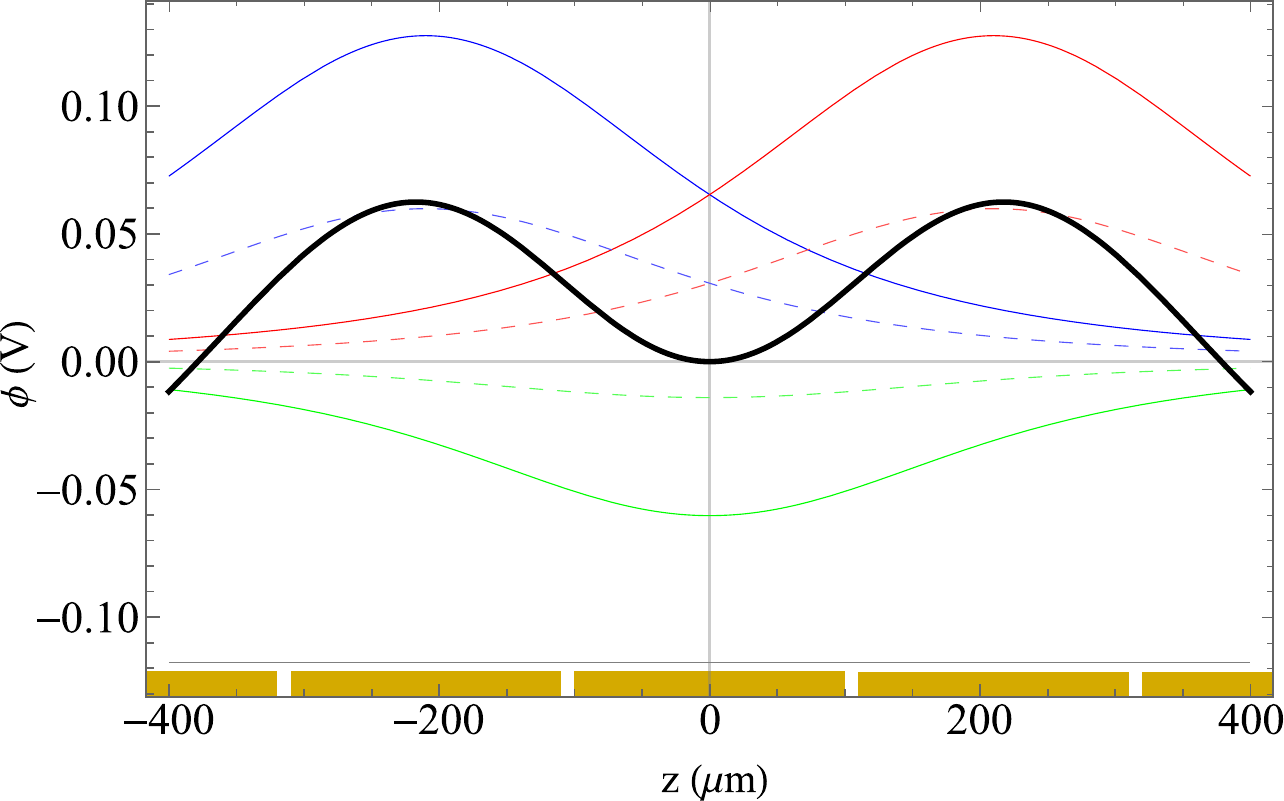}} \hfill
\caption{DC potential along x- and z-axis of the ion trap. (a) The black dotted line is a DC potential without tilting, while the solid black line is the tilted DC potential created by neighboring DC potentials shown in different colors. (b) Tilted DC potential along the axis of the ion trap. The middle DC electrodes are E06 and E16.}

\label{fig:Tilt}
\end{figure}

\begin{figure}[t]%
\subfloat[]{
\includegraphics*[width=0.5\linewidth]{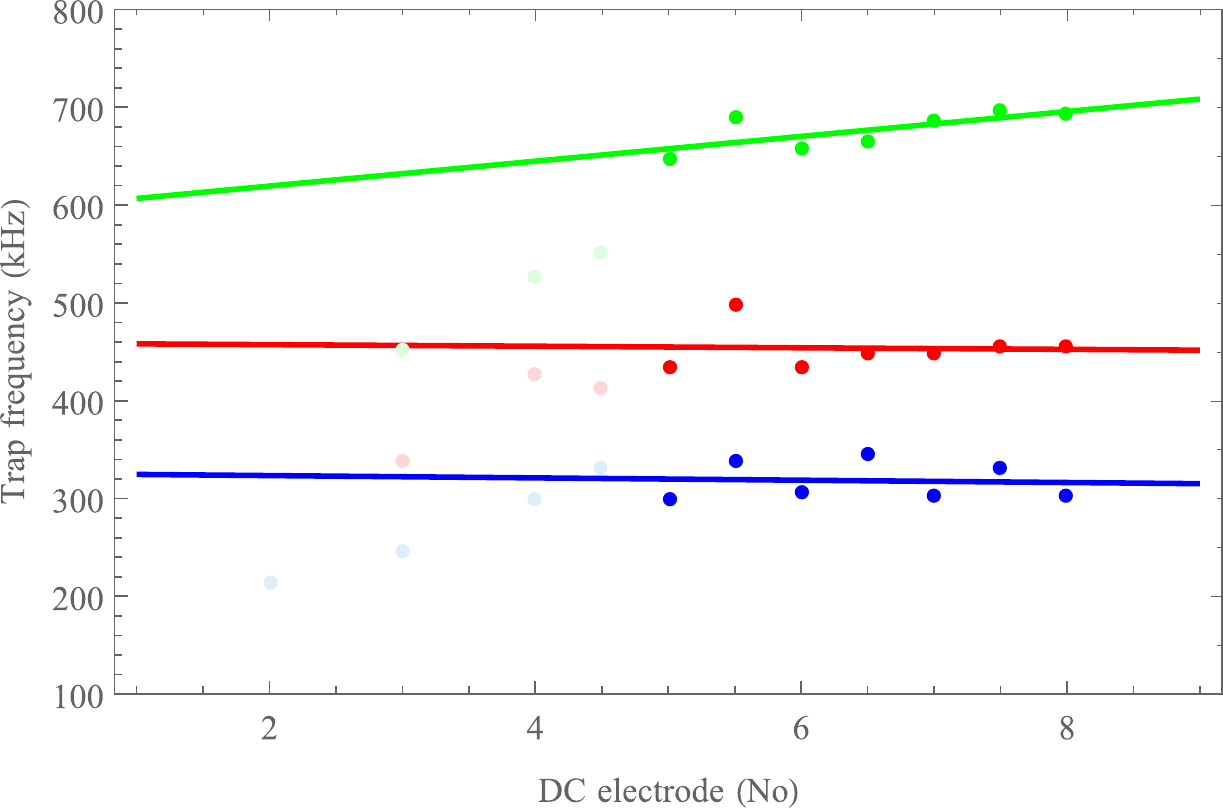}}
\subfloat[]{
\includegraphics*[width=0.5\linewidth]{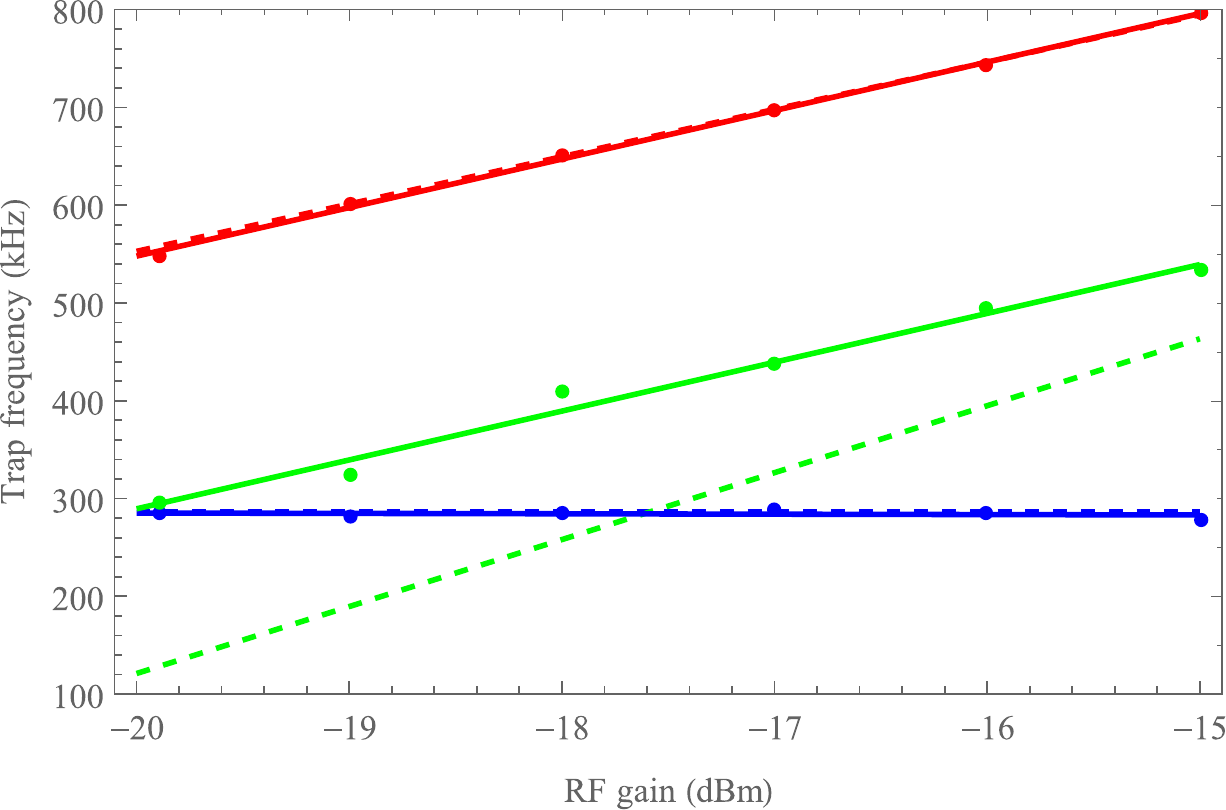}} \hfill

\caption{(a) Trap frequencies along the axis of the ion trap, light dots mark the frequencies for special potentials. (b) Trap frequencies with respect to the applied RF gain. Blue: axial frequency, Green and Red: radial frequencies. Dotted lines are the simulated and solid lines the measured values. The differences result from the tilted axes, which are not taken into account for the calculation of the radial frequencies.
}
\label{fig:TrapFreq}
\end{figure}

In order to effectively cool the trapped ions, we need to tilt the principal axes of the ion so that a single laser can cool all three normal modes. All three Yb laser beams are aligned in parallel to the trap surface such that there is 12\%  projection of its k-vector on the radial trap axis (x and y) and resulting a negligible Doppler cooling efficiency in the radial directions. In order to tilt the principal axes of the ion, we apply DC voltages that are asymmetric along the x-axis (Fig.~\ref{tab:SimUnH}). These DC voltages are extracted following the analytic model for electrostatic fields in planar ion traps \cite{House.2008,Allcock.2010}. To create a non-harmonic potential we set the compensation electrode and one pair of opposing electrodes to different voltages and calculate the voltages for the surrounding electrodes to overlap the minimum of the DC and RF potentials. The tilted DC potential is shown in Fig.~\ref{fig:Tilt}. Since we do not have control over each electrode, due to the shorts on the trap, we need to calculate special fields for electrical shorts to ground for some of the DC electrodes by neighboring electrodes. Note, that even with this constraint, the trap frequencies can be adjusted and measured values fit the theoretical expectation, see Fig. \ref{fig:TrapFreq}.

\subsection{Micromotion compensation}
\label{subsec:micro}
\begin{figure}[t]%
\includegraphics*[width=\linewidth]{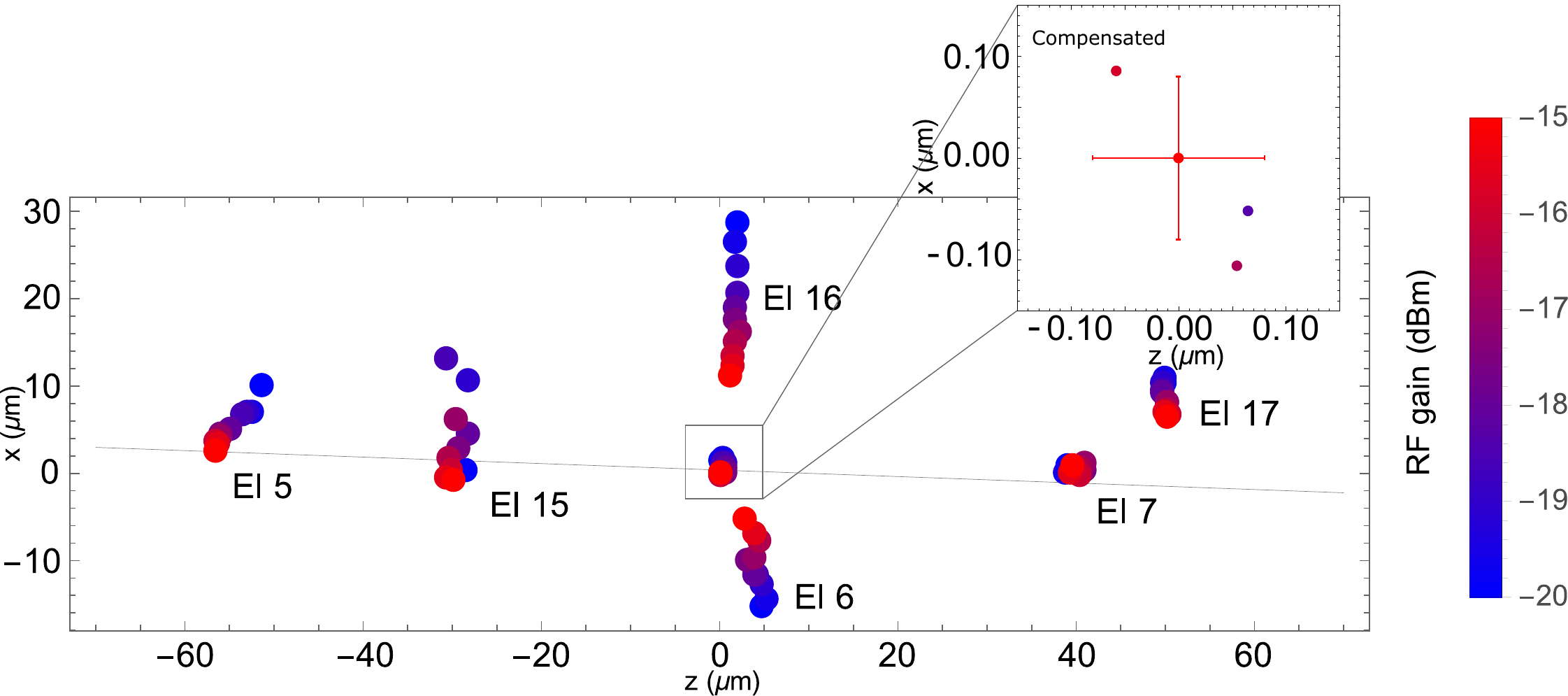}
\caption{Residual stray electric fields after the micromotion compensation. Each dot is showing the position of a single ion at different RF gains. Each group of data is taken with the same set of DC voltages listed in Fig.~\ref{tab:SimUnH} but varied RF gains. The gray line represents the trap axis, extrapolated from the compensated positions over each pair of electrodes. The error bar is the standard error of the Gaussian fit to a single ion image. Voltages used for the micromotion compensation are E05: $0.68\,$V, E06: $-4.5\,$V, E07: $0.68\,$V, E15: $-0.01\,$V, E16: $-3.35\,$V and E17: $-0.51\,$V. The optimized voltage is $-1.0\,$V applied to E06. }

\label{fig:CompAnaly}
\end{figure}

Micromotion compensation is an important and necessary condition for experiments where cold atoms interact with cold trapped ions. In cases, where the center of the DC potential and the RF potential are not superimposed the ions will undergo a fast oscillation at the frequency of the trap drive. This shift has to be determined and corrected by applying the proper corrections to the DC potential. For this, several methods have been applied \cite{Gloger.2015,Ibaraki2011,Berkeland.1998} and the best compensation achieves a residual shift of 0.7$\thinspace\mu$m at respective trap frequencies of $(\omega_{r}, \omega_{z})$ $=2\pi\times(350, 50)\,\text{kHz}$ \cite{Harter.2013}. We vary the RF amplitude and observe on the CCD image how the ion position is changing. In this way, we vary the DC control voltages such that the ion position does no longer change with the variation of the RF amplitude, see Fig.~\ref{fig:CompAnaly}

A systematic measurement of the stray electric fields is done analogues to Ref. \cite{Gloger.2015}. As an example we measure for the case when the ion is over electrodes E06 and E16. We set the surrounding voltages to the simulated set of voltages shown in Fig.~\ref{tab:SimUnH}. The experimental procedure is to vary the RF gain between $-20$ and $-15\,\text{dBm}$ shown at the RF generator and measure the corresponding position of the ion. In each of the following steps we change the voltage of one of the surrounding DCs and repeat the RF variation and position determination. This measurement leads to a 2D plot, where the ion position is always moving towards the RF null for high RF amplitudes. If the compensation is improved there will be less displacement by varying the RF power. One can see, that the simulated values are not optimized. As a conclusion we have to adjust the voltage of electrode E06 to $-1.0\,\text{V}$ to achieve the best compensation. The electrodes sitting right below the ion are important for compensation, while the rest of electrodes are used to shift the position of the ion along the RF node. The residual stray electric field $\varepsilon_{\text{stray}}$ can shift the ion position away from the RF node to position 
\begin{equation}
	\textbf{r}_{\varepsilon} \approx \frac{q}{m} \sum_{k} \frac{\varepsilon_{\text{stray},k}}{\omega^2_k} \hat{\textbf{k}}, 
\end{equation}

where $\textbf{r}_{\varepsilon}$ is the average position of the ion in $\hat{\textbf{k}}=(x, y, z)$ direction, $q$ the charge of the ion, $m$ the mass of the ion and $\omega_k$ the secular frequencies. We can measure the $\textbf{r}_{\varepsilon}$ from a single ion image and therefore extract the residual stray electric field $\varepsilon_{\text{stray}}$. After the 2D compensation we estimate a residual electric stray field uncertainty of $(\textDelta\varepsilon_{\text{stray},x}, \textDelta\varepsilon_{\text{stray},z})=(0.90, 0.30)$ V$/$m with the corresponding trap frequencies of $(\omega_{z}, \omega_{x})$ $=2\pi\times(305, 435)\,\text{kHz}$. The residual electric stray field uncertainty measured in our ion trap are sufficient enough for a first atom-ion interaction study but in the future we will extend the 2D micromotion compensation to a full 3D by using parametric resonance excitation induced by potential modulation \cite{Ibaraki2011}.

% CONCLUSION %%%%%%%%%%%%%%%%%%%%%%%%%%%%%%%%%%%%%%%%%%%%%%%%%%%%%%%%%%%%%%%%%%%%%%%%
%%%%%%%%%%%%%%%%%%%%%%%%%%%%%%%%%%%%%%%%%%%%%%%%%%%%%%%%%%%%%%%%%%%%%%%%%%%%%%%%%%
\section{Conclusion and outlook}
\label{conclusion}
We have introduced a hybrid atom-ion experimental setup to trap the Yb$^{+}$ ions and the $^{87}$Rb atoms to investigate atom-ion interaction. Our setup is a compact planar Paul trap with segmented DC electrodes, featuring the exact positioning of single ions. We characterized the trap performance including minimization of micromotion arising from stray electric fields. Comparing the residual electric stray fields to literature the micromotion compensation should be good enough for atom-ion interactions. Controlled movement of the ions in the axial direction of the trap for about $1.6\,$mm is also accomplished. The next step in our experiment will be trapping $^{87}$Rb atoms before we overlap them with the ion crystals in order to investigate atom-ion collision dynamics.

\begin{acknowledgement}
This work was supported by the EU-STREP EQuaM and the SFB/TR49. We thank Johannes Denschlag (Univ. of Ulm), Hartmut H{\"a}ffner (Univ. of California Berkeley) for support and Arezoo Mokhberi for helpful discussions. A. Bahrami and M. M\"uller contributed equally to this work.
\end{acknowledgement}

% Use the following code if you wish to generate your bibliography with BibTeX;
% replace the string "pss_demo" below with the name(s) of
% the BibTeX data base(s) you want to use.
% The resulting bibliography-output (the content of the .bbl file)
% must be pasted back into this file before submission.
% Please also include your BibTeX data base file(s) in your submission
% so that we can re-run BibTeX if necessary.
%
\bibliographystyle{pss}
\bibliography{pss_demo}
%
% Replace the following example bibliography with your references
% before submission:

%\begin{thebibliography}{[1]}

%\bibitem{bib2}%
 %F.~Examplename and  I.\,E. Anotherauthorname,
% Phys. Status Solidi A \textbf{1}, 111 (2050).

%\bibitem{bib3}%
% A.\,N.~Earlyview, E.\,X.~Ample, and Y.\,A.~Author,
% Phys. Status Solidi A, DOI 10.1002/pssr.205001234  (2050).

%\bibitem{bib4}%
 %A.~Firstauthorname,  B.~Secondauthorname,  and
 % C.~Thirdauthorname,
%Here Goes the Title of the Book (Publisher, City, year), p.\,111.

%\bibitem{bib5}%
% A.~Firsteditorname,  B.~Secondeditorname,  and
 % C.~Thirdeditorname (eds.),
%Here Goes the Title of the Edited Book (Wiley-VCH, Berlin, 2050), p.\,111.

%\bibitem{bib6}%
% D.~Contributorname,
% in: The Title of the Book, edited by The Name of the Editors, Followed by
%  the Title of the Series of Books (Publisher, City, year), chap.~1.

%\bibitem{bib7}%
% A.~Lastbutnotleastname,
% Proceedings 1st Dummy Conference on Citation Formatting, City,
%  Country, Part A (Publisher, City, year),  pp.\,1--11.

%\end{thebibliography}

\newpage

\section*{Graphical Table of Contents\\}
GTOC image:
\begin{figure}[h]%
\includegraphics[width=4cm,height=4cm]{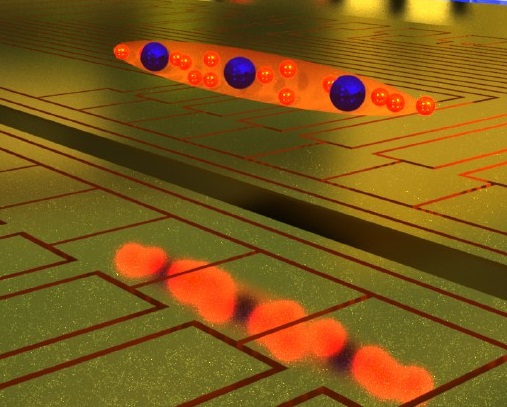}
\caption*{%
We introduce a new micro-fabricated planar ion trap combined with an integrated atom trap into a single chip. Trapped single ions can be coupled to a small neutral atom cloud enabling us to investigate atom-ion interactions. This includes the tunneling dynamics of a bosonic Josephson junction and spin dynamics of ions inside polarized neutral atoms.}
\label{GTOC}
\end{figure}

\end{document}